# Decomposition Sampling Applied to Parallelization of Metropolis–Hastings

**Jonas Hallgren** · **Timo Koski**

**Abstract** This paper presents an algorithm for sampling random variables that allows to separation of the sampling process into subproblems by dividing the sample space into overlapping parts. The subproblems can be solved independently of each other and are thus well suited for parallelization. Furthermore, on each of these subproblems it is possible to use distinct and independent sampling methods. In other words, specific samplers can be designed for specific parts of the sample space. The algorithms are demonstrated on a particle marginal Metropolis–Hastings sampler applied to calibration of a volatility model and two toy examples. Significant speedup and decrease of total variation is observed in experiments.

**Keywords** Parallel Computing · Metropolis–Hastings · MCMC · Stochastic volatility.

## 1 Introduction

The Metropolis–Hastings (MH) algorithm, by Metropolis et al (1953) and Hastings (1970), generates a Markov chain which after reaching stationarity produces samples from a specified distribution. A drawback of the otherwise versatile MH algorithm is that it is unclear how to execute it in parallel since every new iteration of the algorithm depends on the preceding. The same is true for other Markov chain Monte Carlo (MCMC) methods.

A straightforward way to parallelize the algorithm is to simply run many chains in parallel and then, after removing burn-in, combine them into one single chain. This was done in Rosenthal (2000) and more than ten years later in a modern setting by Henriksen et al (2012). However, combining the samples to a single chain is not equivalent to running one long chain. For large state spaces it is unlikely that good starting values are found and thus that the combined chain will have converged.

This paper suggests Decomposition (DC) sampling, a parallel algorihtm for exactly sampling random variables by dividing the sample space into parts. This division allows a faster and more efficient exploration of the sample space without compromising the integrity of the samples. The strength of the algorithm lies rather in these properties than in the ability to do massive parallelization.

Finding a good divison is not trivial and results does not necessarily continue to improve as the number of parts increase.

We therefore think that there are two important main uses for the algorithm: Distributed computing and small specialized systems. In distributed computing there are a relatively small number of computers but each computer is powerful and might be equipped with a Graphical Processing Units (GPU) with thousands of processors. Small specialied systems which solve similar problems repeteadly and are limited in hardware are also potential candidates. An example is planetary rovers which use particle filtering methods Dearden and

Funding for this research was provided by the Swedish Research Council (Grant Number 2009-5834).

J. Hallgren
Department of Mathematics, KTH, Royal Institute of Technology, Stockholm SE-100 44
Tel.: +46-8-7906619
E-mail: jonas@math.kth.se

Timo Koski
Department of Mathematics, KTH, Royal Institute of Technology,

Clancy (2002). Another use is for small desktop and laptop computers which typically have a small number of processing units.

The speed of convergence is demonstrated on a discrete sample space where a a speedup of 10,000 times is obtained.

The main application of the paper is a calibration of a stochastic volatility model using Particle Marginal Metropolis–Hastings sampler (PMMH), Andrieu et al (2010). Here it is seen that the autocorrelation between samples decreases faster for the DC-sampling than for standard MH.

The computing by Henriksen et al (2012) is done on GPU hardware and is also an implementation of the PMMH. This approach can provide an almost linear speedup, which is ideal, when the convergence rate is fast. However when the convergence rate is slow the result may be a combination of chains of which none have reached stationarity. Their parallelization of the particle filter on the other hand does not suffer at all from this.

A review of a few approaches to parallelization reveals that they are often limited to different special cases of MCMC. The Bayesian model learning by Corander et al (2006) is done on a finite state space. Jacob et al (2011) presents a method that does parallelize a special case, the independent MH sampler. Altekar et al (2004) implements a parallelization of Metropolis coupled MCMC, $(MC)^3$. This approach deals with the fact that $(MC)^3$ is slower than standard MCMC and while the method yields a nice speedup it is, like Jacob et al (2011) and Corander et al (2006), application specific. The population Monte Carlo method introduced by Cappé et al (2004) does not suffer from these drawbacks, has many advantages, and is not limited to MCMC. Lee et al (2010) implements population Monte Carlo and other algorithms in a GPU framework providing significant speedups. Suchard et al (2010) discusses technical aspects of GPU programming and provide guidelines on how to implement GPU simulations in an MCMC setting.

Recent developments of parallel methods include methods similar to the one here proposed. VanDerwerken and Schmidler (2013) suggest a partitioning and weight estimating scheme that partitions the state space into nonoverlapping sets.

Minsker et al (2014), Neiswanger et al (2013), Scott et al (2013), and Xu et al (2014), all use partitioning but of the data instead of the state space.

Despite the similar ideas of dividing in space instead of time there is an important distinction between the two approaches: Decomposition sampling and the approach proposed by VanDerwerken and Schmidler (2013) partitions the state space while the other algorihtms partitions the data. The methods can be combined but they solve different problems. Decomposition sampling does nothing to reduce the size of the data; instead the complexity of the sampling procedure is reduced. Alternatives to partitioning the data is for instance subsampling, see e.g. Quiroz et al (2014) or Bardenet et al (2015).

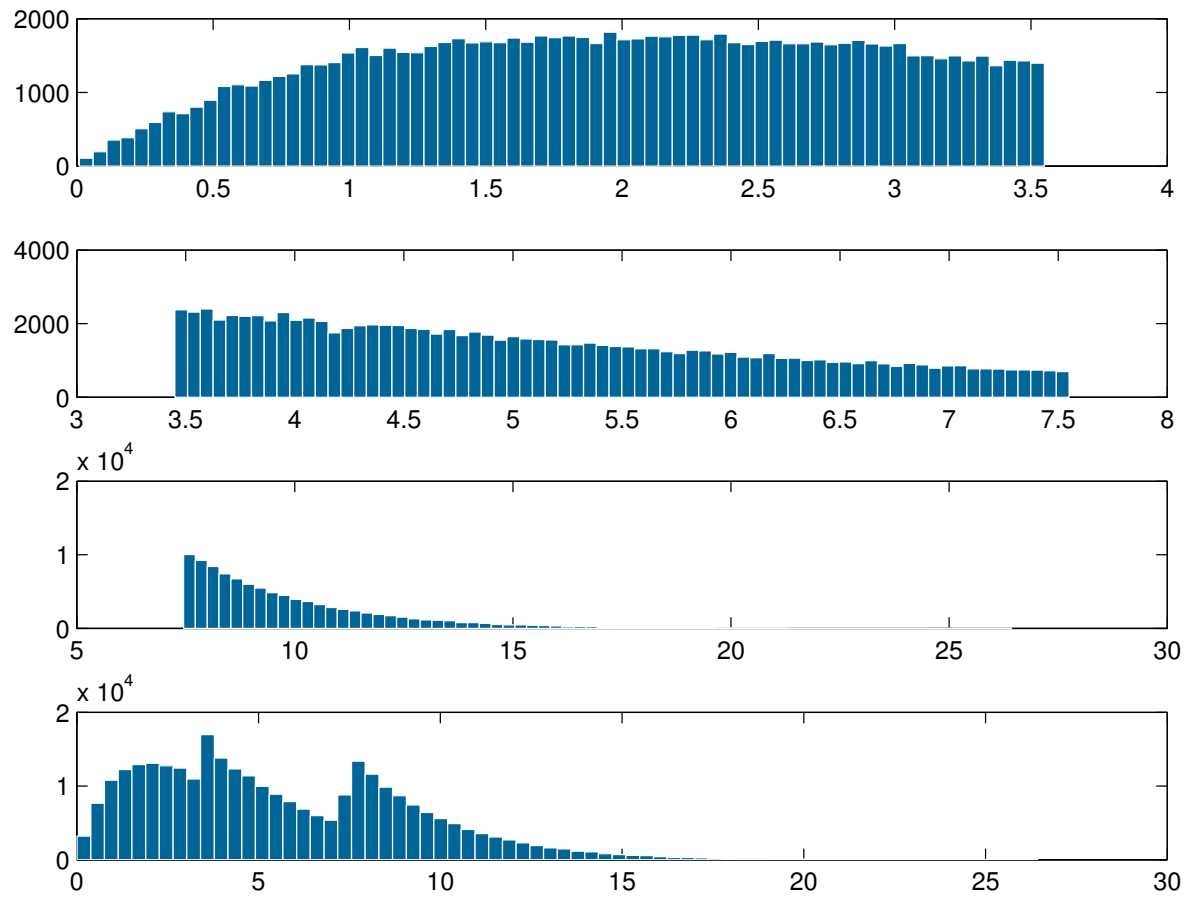

**Fig. 1** Unprocessed samples from the gamma distribution

Other non MCMC methods for sampling random variables such as importance sampling, rejection sampling, or inverse transform sampling, all of which are presented by Robert and Casella (2004), are often easy to parallelize simply because they are parallel in their nature; no sample is drawn in relation to another. It is important to stress that sometimes the most efficient speedup is not attained by parallelizing the full sequential procedure; but rather the comprising subprocesses as in Henriksen et al (2012). However, more than being able to parallelize, the method presented can significantly speed up convergence.

The main algorithm presented is based on a simple idea: decompose the sample space into several parts, subsets, then sample on these subsets independently of each other. If the probability of landing in one subset is higher than in another, discard some of the samples in the less likely subset. If the probabilities to land in the subsets are unknown, we can obtain them by evaluating integrals on the intersections of the subsets. This requires the subsets to be intersected. Furthermore, it allows the different samplers to explore the full space independently of each other. This in contrast to for example the Parallel Wang–Landau by Bornn et al (2013) where the chains need to interact making it difficult to run independent batches. An important point is that while this paper the method is mainly applied to MCMC methods it is not limited to those and there may be advantages in applying the algorithm to other methods.

The paper is divided into four sections. The first is this introduction that presents some background and terminology. In Section 2 we introduce necessary nomenclature and objects, and present the three main algorithms.

1. The first algorithm is the main one; it merges samples drawn in subsets into a single sample drawn from the full space.
2. The second algorithm produces expectations evaluated on the full space.
3. The third algorithm provides the first two algorithms with the proportion between the subsets if the proportion is unknown.

In this section there are also a few demonstrating examples and a proposed algorithm for automatically selecting the cover. In Section 3 three examples are evaluated. A stochastic volatility model is calibrated using PMMH Andrieu et al (2010). Moreover, two toy examples are studied: a Markov chain with a small discrete space especially well suited for the method. Finally the example with a Poisson distribution from Hastings (1970) is replicated. The efficiency of the method is examined and the results are presented.

Subsection 3.5 discusses potential improvements.

The proofs in the paper and the theoretical results are, with one exception, given in the appendix.

## 2 Decomposition Sampling

Let $X$ be a random variable on the probability space $(\mathsf{S}, \mathcal{B}(\mathsf{S}), \mathbb{P})$ where $\mathcal{B}(\mathsf{S})$ is the Borel sigma-algebra of $\mathsf{S}$. Typically $\mathsf{S}$ is a subset of a high-dimensional real space or integer lattice. Assume that there exist a specific cover $C = \{C_1, C_2, \ldots, C_W\}$ of $\mathsf{S}$ and define $\Delta_{j,k} = C_j \cap C_k$, and $\Delta_j = C_j \cap C_{j+1}$ where $\Delta_0 = \Delta_W = \varnothing$. That is, assume that there exists a cover such that every element of the cover shares two distinct subsets of itself with the previous and following element in the cover. This cover shall be called a *linked cover* of $\mathsf{S}$. If a linked cover contains only elements and intersections such that the probability that $X$ is in them is strictly positive with respect to $\mathbb{P}$, we say that $C$ is a *saturated linked cover*. The task is to sample a random variable taking values in $\mathsf{S}$. When $\mathsf{S}$ is the positive real line, an example of a linked cover is

$$C_1 = [0, 3.55], \quad C_2 = [3.45, 7.55], \quad C_3 = [7.45, \infty], \tag{1}$$

implying that the intersection $\Delta_1$ is given by $[3.45, 3.55]$. Assume that samples from a gamma distribution are desired. The linked cover from (1) is used. Sampling from each of the subsets in $C$ yields the results presented in the first three subplots in Figure 1. In the fourth subplot the three samples are combined into one in a histogram. Clearly, there is an abundance of samples in $C_2$ and in $C_3$ that would not be there if this truly was a sample from the gamma distribution. This is where the intersections come into play. An integral evaluated on the intersection $\Delta_1$ using samples from $C_1$ should, needless to say, have the same value as an integral evaluated on $\Delta_1$ using samples from $C_2$. This allows us, by looking at $\frac{\mathbb{P}(X \in C_1)}{\mathbb{P}(X \in C_2)}$, to find out how many of the abundant samples in the tail to discard. The samples from Figure 1 are processed through the algorithm throwing out the abundant samples, producing the result in Figure 2.

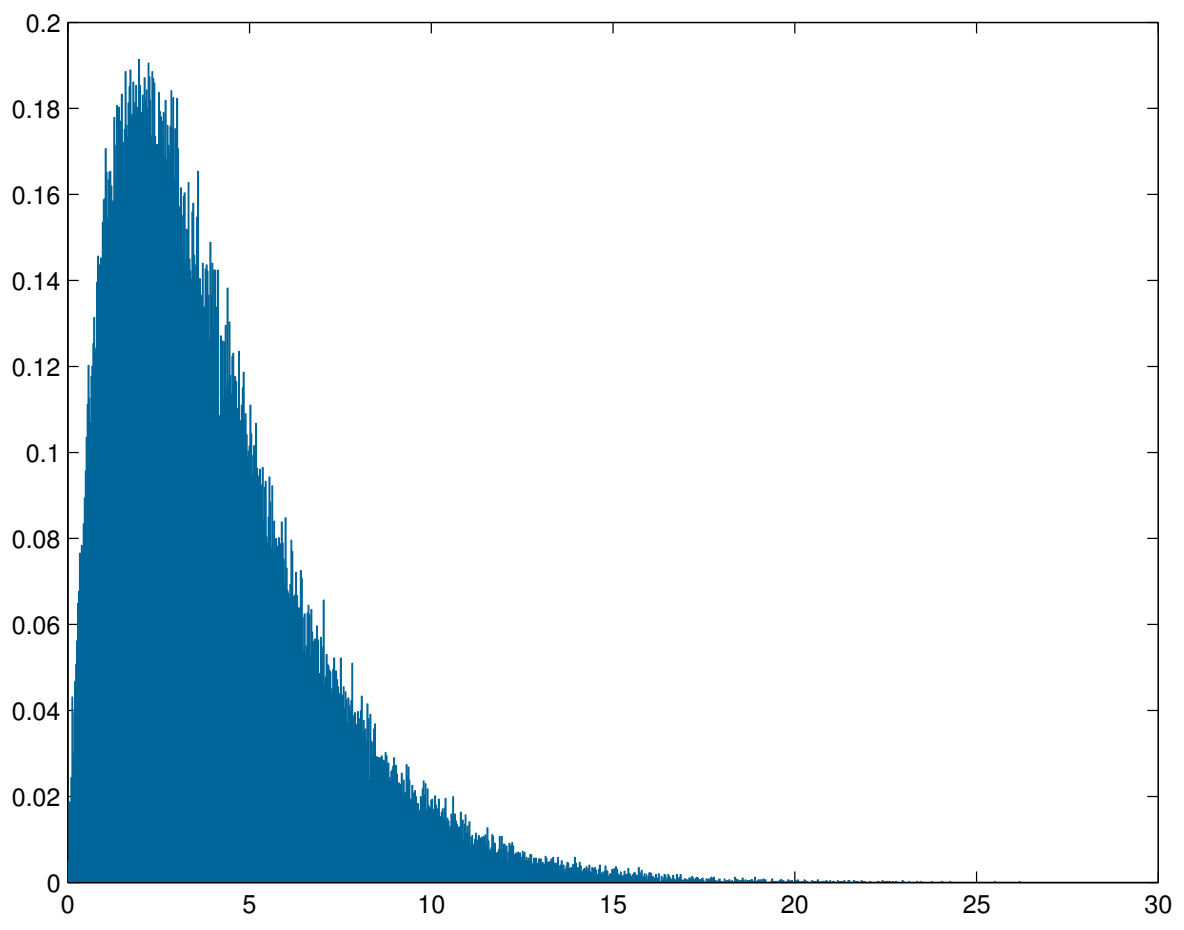

**Fig. 2** Processed samples from Figure 1

An example of a discrete space linked cover is seen in Figure 3. There is a particularly vicious transition probability where it is very difficult to reach state 4 from state 3 and very easy to reach it from state 5. This means that to simulate a Markov chain and start in state $1, 2, 3$ or $4$, it is very unlikely to reach $5, 6$, or $7$ in a small number of steps. Indeed, simulations indicate that starting in state 4, requires copious amounts of steps to reach state 5. Having said that, by splitting the space two parts, $C_1$ and $C_2$, allows a simulation exploring the full space in a small number of steps. In addition it can be done in parallel. More details and numerical results are found in Sections 3 and 3.5.

The integer $W$ will be used throughout the paper to denote the number of *agents* used. An agent is a machine that can execute an algorithm. In practice a computer typically produces one agent for every core of the CPU; although, an agent could also be a cluster of computers and a single-core processor could simulate several agents. Let the backslash symbol denote the set difference. The basis for the main algorithm rests on the following identities.

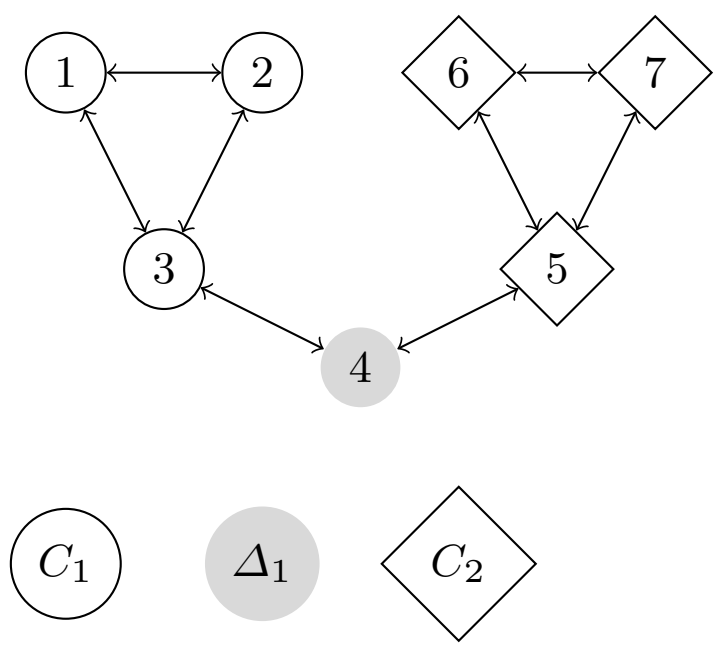

**Fig. 3** Discrete space linked cover

Let $C = \{C_1, \dots, C_W\}$ be a saturated linked cover of $\mathsf{S}$. Let $A \in \mathcal{B}(\mathsf{S})$. Hereafter, let

$$\pi(A) = \mathbb{P}(X \in A)$$

and $\pi(A \mid B) = \mathbb{P}(X \in A \mid X \in B)$. Then since $A$ can be written as a union of disjoint sets, $\bigcup_{j=1}^{W} \{A \setminus \Delta_{1:j-1}\} \cap C_j$, the probability of event $A$ can be written as the sum

$$\begin{aligned}\pi(A) &= \sum_{j=1}^{W} \pi(\{A \setminus \Delta_{1:j-1}\} \cap C_j) \\ &= \sum_{j=1}^{W} \pi(A \setminus \Delta_{1:j-1} \mid C_j)\pi(C_j),\end{aligned} \tag{2}$$

where $\Delta_{1:j-1} = \bigcup_{k=1}^{j-1} \Delta_{k,j}$. Furthermore, when $j > 1$, note that

$$\pi(\Delta_{j-1}) = \pi(C_j \cap \Delta_{j-1}) = \pi(C_{j-1} \cap \Delta_{j-1})$$

so

$$\begin{aligned}\pi(\Delta_{j-1}) &= \pi(\Delta_{j-1} \mid C_j)\pi(C_j) \\ &= \pi(\Delta_{j-1} \mid C_{j-1})\pi(C_{j-1}).\end{aligned}$$

Dividing by $\pi(\Delta_{j-1} \mid C_j)$ gives the relation

$$\pi(C_j) = \pi(C_{j-1}) \frac{\pi(\Delta_{j-1} \mid C_{j-1})}{\pi(\Delta_{j-1} \mid C_j)}. \tag{3}$$

That is, the probability of $X$ being in $C_j$ is equal to that of $X$ being in $C_{j-1}$ multiplied by the relative size of the probability that $X$ is in the intersection of the two sets.

## 2.1 Merging

Let $C = \{C_1, \dots, C_W\}$ be a saturated linked cover of $\mathsf{S}$. Assume that there are available samples of $X$ from the different sets in $C$. Below these independent samples are merged into one single sample distributed as $X$. The following algorithm will generate samples $\xi_k^{C_j}$ at each iteration $k$. The samples will be drawn from a *proposal density* $q_k^{C_j}$ which may depend on $\xi_{1:k-1}^{C_j}$. That is, each sample from $C_1, C_2, \dots, C_W$ may be generated by different densities. In the current setting they are allowed to depend on their individual history only. Interactions between the samplers $q_k^{C_j}$ might be possible but would make batch processing problematic. An MH sampler would have $q_k^{C_j}(\cdot) = r^{C_j}(\cdot \mid \xi_{k-1}^{C_j})$, where $r$ is a transition density. A sequence $\eta = \{\eta_k\}_{k \geq 1}$, taking values in the full space $\mathsf{S}$ is generated by merging the samples of $\xi$. This is precisely what Algorithm 1 does.

**Algorithm 1** Decomposition sampling

> **for** $k = 1, \dots, M$ **do**
>   **for** $j = 1 \dots, W$ **do**
>     Sample: $\xi_k^{C_j} \sim q_k^{C_j}(\cdot)$.
>     Downsampling: with probability
>
> $$\frac{\pi(C_j)}{\max_r \pi(C_r)} \mathbb{I}(\xi_k^{C_j} \notin \Delta_{1:j-1}), \text{ put } \tilde{\eta} \leftarrow \{\tilde{\eta}, \xi_k^{C_j}\}.$$
>
>   **end for**
>   Randomize the order of the samples $\tilde{\eta}$.
>   Put them in the main sequence: $\eta \leftarrow \{\eta, \tilde{\eta}\}$.
> **end for**

For a finite $M$ it is possible to first sample the trajectories $\xi$ in batch and then, given $\xi$, merge them into a single trajectory $\eta$. Batch sampling is typically preferred to running the algorithm online, but both approaches permit the different samples of $\xi$ to be drawn in parallel.

It is possible to use a weighted sample instead. Assuming that $M$ is finite allows the order of the for loops to swap order which reduces the algorithm to drawing a weighted sample.

That would give a weighted sample which can produce samples by first drawing an index $J$ from $\pi(C_j \setminus \Delta_{1:j-1})$ and then uniformly select a sample from

$$\{\xi_k^{C_J \setminus \Delta_{1:J-1}}\}_{k=1}^{M}.$$

It is also possible to reuse the discarded samples $\xi_k^{\Delta_{1:j-1} \mid C_j}$. Let $C_s$ be the set where the discarded sample would be allowed. Then separate the intersected samples from the other ones. Now, given that a sample was selected from $s$, sample uniformly from $\xi_k^{\Delta_s \mid C_s}$ with probability $\pi(\Delta_s \mid C_s)$ and otherwise uniformly from $\xi_k^{C_s \setminus \Delta_{1:s}}$.

Given the main algorithm we will introduce two other algorithms that are of practical use. Algorithm 2 uses the weighted sample to evaluate expectations of the type $\mathbb{E}[h(X)]$ where $h \in L^1[\pi(\cdot \mid C_j)]$ for every $j$. Algorithm 3 is used to supply the other two algorithms with the proportionality between the subsets.

**Algorithm 2** Evaluating Expectations

**for** $j = 1, \ldots, W$ **do**
  $\{\xi_k^{C_j}\}_{k=1}^M \sim \pi(\cdot \mid C_j)$. ▷ Sample a sequence of i.i.d. random variables
**end for**
**for** $k = 1, \ldots, M$ **do**
  $\eta_k \leftarrow \sum_{j=1}^W h(\xi_k^{C_j})\mathbb{I}(\{\xi_k^{C_j} \notin \Delta_{1:j-1}\})\pi(C_j \setminus \Delta_{1:j-1})$.
**end for**

**Algorithm 3** Estimating $\pi(C_j)$

**for** $j = 1, \ldots, W$ **do**
  Sample an i.i.d. sequence $\{\xi_k^{C_j}\}_{k=1}^M \sim \pi(\cdot \mid C_j)$
  If $j = 1$ put $\pi(C_j)^{(M)} \propto 1$. If $j > 1$ put

$$\hat{\pi}^M(C_j) \leftarrow \hat{\pi}^M(C_{j-1}) \frac{\sum_{k=1}^M \mathbb{I}(\xi_k^{C_{j-1}} \in \Delta_{j-1})}{\sum_{k=1}^M \mathbb{I}(\xi_k^{C_j} \in \Delta_{j-1})}. \quad (4)$$

**end for**
Normalize, for $j = 1, \ldots, W$,

$$\pi^M(C_j) \leftarrow \frac{\hat{\pi}^M(C_j)}{\sum_{s=1}^W \hat{\pi}^M(C_s)\{1 - M^{-1}\sum_{k=1}^M \mathbb{I}(\xi_k^{C_s} \in \Delta_{s-1})\}}.$$

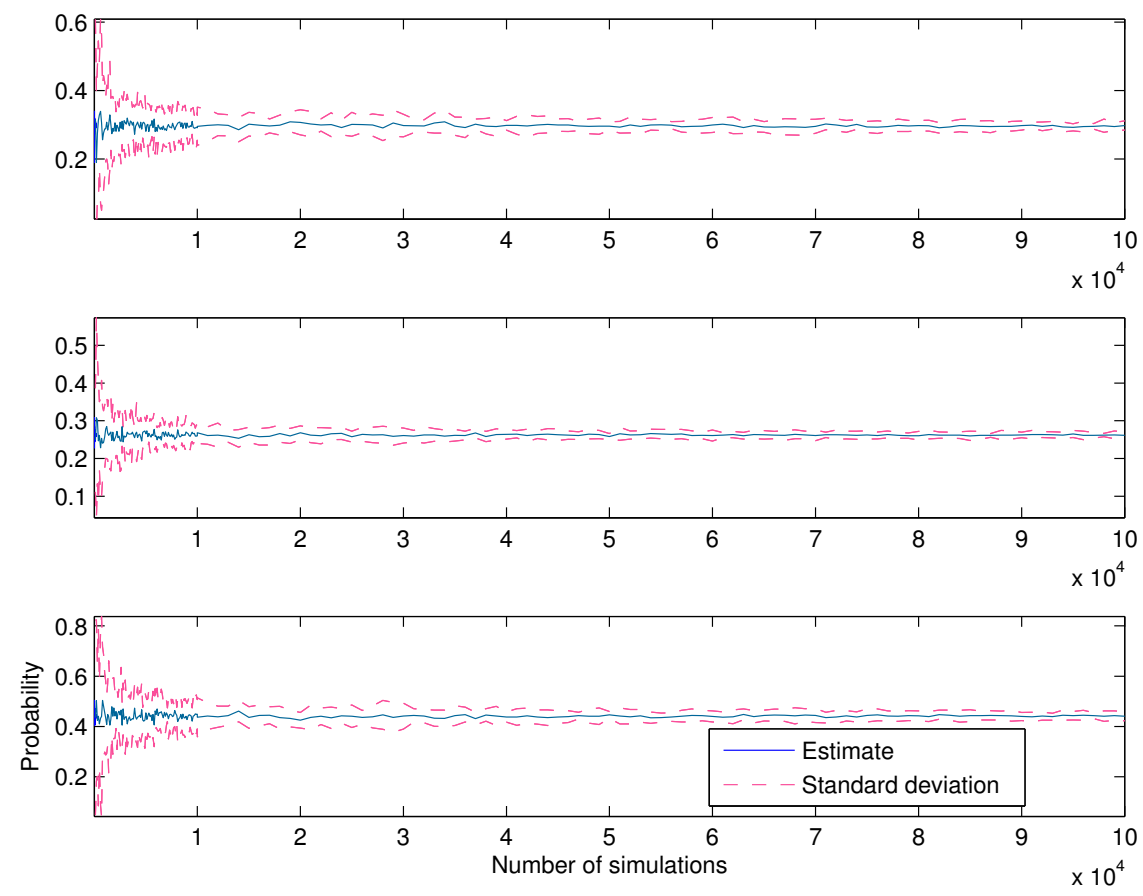

**Fig. 4** Convergence of the estimated $\pi(C_j)$ for a gamma distribution. The three plots are in order for $j = 1, 2, 3$.

*2.1.1 Probability of Failure*

If the algorithm does not produce at least one sample from every intersection for every $C_j$ it has failed. The probability of failure is the probability that for least one of the $C_j$s there is a $\Delta_j$ or $\Delta_{j-1}$ with zero samples. In total there are $2(W-1)$ such regions. This since each $C_j$ has two $\Delta$-regions except for the first and the last which have one each. The probability of failing in the worst section is

$$\max\Big\{ \max_{j \in 1,\ldots,W-1} \mathbb{P}(X \notin \Delta_j \mid X \in C_j), \max_{j \in 2,\ldots,W} \mathbb{P}(X \notin \Delta_j \mid X \in C_{j-1})\Big\},$$

denoted $p_{\text{worst}}$. The probability of obtaining at least one sample (preventing failure) in the worst section in $M$ trials is $1 - p_{\text{worst}}^M$. The probability that the algorithm fails is the probability of the complement that it does not fail on any of these sections. Thus we compute the probability of the algorithm failing in $M$ trials as

$$\begin{aligned}\mathbb{P}(\text{Fail}) &= 1 - \mathbb{P}(\text{Not failing in any section})\\ &\le 1 - (1 - \mathbb{P}(\text{Failing in the worst section}))^{2(W-1)}\\ &= 1 - (1 - p_{\text{worst}}^M)^{2(W-1)}.\end{aligned}$$

From this it is clear why it is required that the linked cover is saturated; otherwise, $p_{\text{worst}}$ is equal to zero and the algorithm fails with probability 1. As $M$ grow large, the probability of failure fades away. Experiments show that for small $M$ the algorithm fails. In the creation of the convergence plot, Figure 4, the algorithm never failed for $M$ larger than 100 but often when it was equal to ten.

Convergence results in form of total variation, and almost sure convergence are given in Appendix A. For a concise overview of these concepts and alternatives such as a Central Limit Theorem see Nummelin (2002).

## 2.2 Subspace sampling

The problem of sampling on the subspaces instead of the full space poses no major difficulties since

$$\pi(A \mid C_j) = \frac{\pi(A, C_j)}{\pi(C_j)} \propto \pi(A, C_j) = \pi(A)\mathbb{I}(A \in C_j).$$

Thus, for random walk MH-sampling, the procedure needs to be adjusted only by evaluating the indicator function. If the proposals are not generated by a random walk but by a proposal over the full space the acceptance ratio will decline; however, typically there is no good such proposal when decomposition sampling is of interest. A simple such proposal distribution will often be possible to adapt to a subcover.

Rejection sampling requires

$$\pi(A) < Mq(A);$$

since $\pi(A \mid C_j) \leq \pi(A)$ the procedure can be used without adjustment.

Importance sampling is, like MH-sampling, independent of the normalizing constant and needs no adjustment.

In Gibbs sampling, Geman and Geman (1984), where samples are generated from marginal distributions a rejection sampling approach would have to be used. If the variables are uncorrelated the acceptance rate will be low and it is likely that it is better to do standard Gibbs sampling. However, for highly correlated variables it is more likely that a sample will be generated that lies in the cover.

### 2.3 Selection of the Cover

The selection of the cover is important. By studying the variance of the $\pi$ estimate we find that it is advantageous if the intersections are chosen such that the intersections have an approximately equal probability. A natural way to select the cover is therefore

$$C_1 = \{x : 0 \leq \mathbb{P}(X \leq x) \leq \tfrac{1}{W} + \frac{\delta}{2}\},$$

$$C_j = \{x : \tfrac{j-1}{W} - \frac{\delta}{2} \leq \mathbb{P}(X \leq x) \leq \tfrac{j}{W} + \frac{\delta}{2}\},$$

$$C_W = \{x : \frac{W-1}{W} - \frac{\delta}{2} \leq \mathbb{P}(X \leq x) \leq 1\},$$

where $\delta$ is a design parameter. Set $x_j^+ = \sup_x x \in C_j$ and $x_{j+1}^- = \inf_x x \in C_{j+1}$. Then $\mathbb{P}(x_{j+1}^- \leq X \leq x_j^+) = \mathbb{P}(X \in \Delta_j) = \frac{j}{W} + \frac{\delta}{2} - \frac{j}{W} + \frac{\delta}{2} = \delta$. This is a desirable property since it reduces variance as can be seen in the proof of Proposition 2. When the algorithm is utilized, these probabilities will be unknown. In the Bayesian setting a guess could be obtained from the priors. When no prior information is available one could either assume uniformity or possibly run a few initial simulations to get a rough idea of the appearance. From this an algorithm materializes: first run trial simulations then estimate the quantities needed to generate the cover above.

**Algorithm 4** Estimating the cover

Sample an i.i.d. sequence $\{\xi_k^{\text{Pilot}}\}_{k=1}^{M_{\text{Pilot}}} \sim \pi(\cdot)$
$\mathbb{W} \leftarrow$ DisperseCover($W$, Dimensions)
**for** $j \in$ Dimensions **do**
  $B \leftarrow$ Quantiles($\xi^{\text{Pilot}}(j)$)
**end for**
$\hat{C} \leftarrow$ MergeCover($\delta, B, W, \mathsf{S}$, Dimensions)

The called functions are available as MATLAB implementations in the digital attachment but are briefly described below.

Typically $W$ is given as an integer so a suitable number of parts for each dimension is needed. That is, we need to divide an $n$-dimensional hyperrectangle into $W$ distinct parts which is equivalent to finding a factorization of $W$ with $n$ factors. The function DisperseCover($W$,$n$) does that. If $W$ is a power of two greater than or equal to $2^n$ it is always possible to find a factorization giving an equal amount of parts in each dimension. In other cases there might be more parts in some dimensions and in the case when $W$ is prime all parts must lie in one dimension. DisperseCover only computes the number of parts in each dimension, not the size.

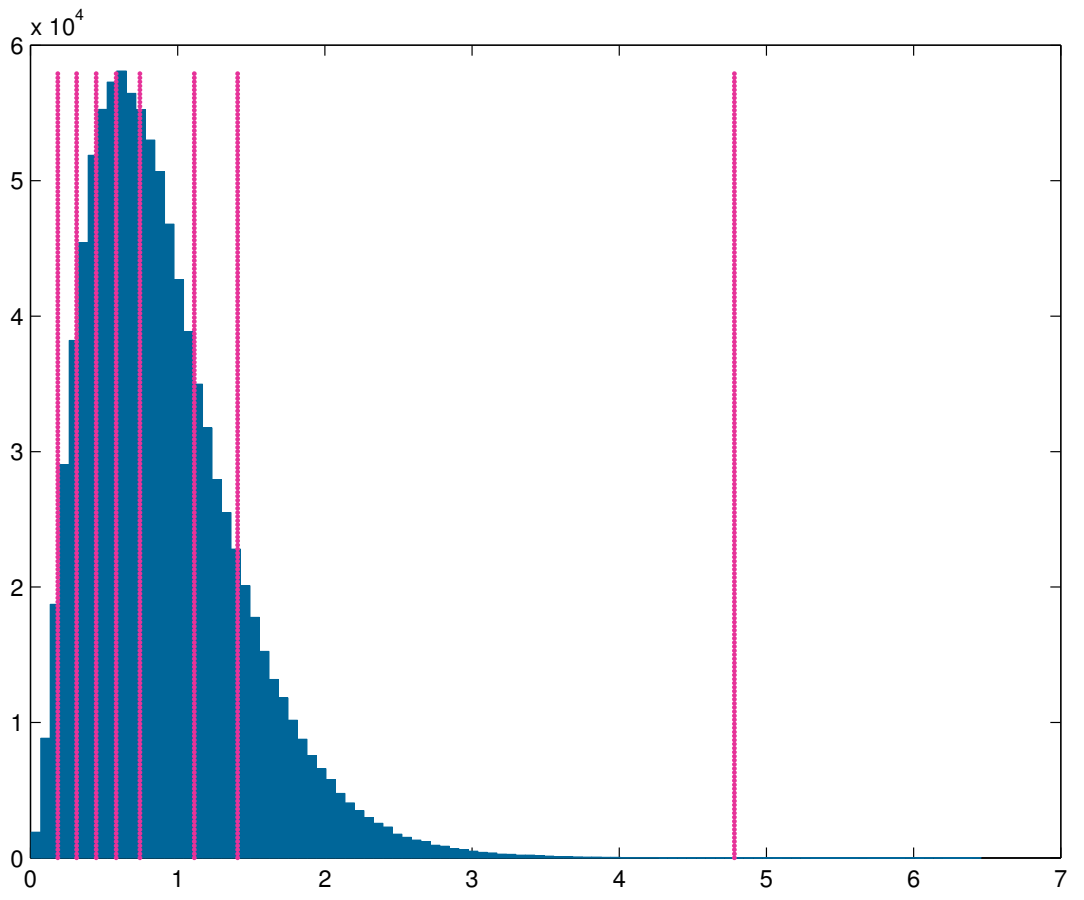

**Fig. 5** Example of an estimated cover for the gamma distribution

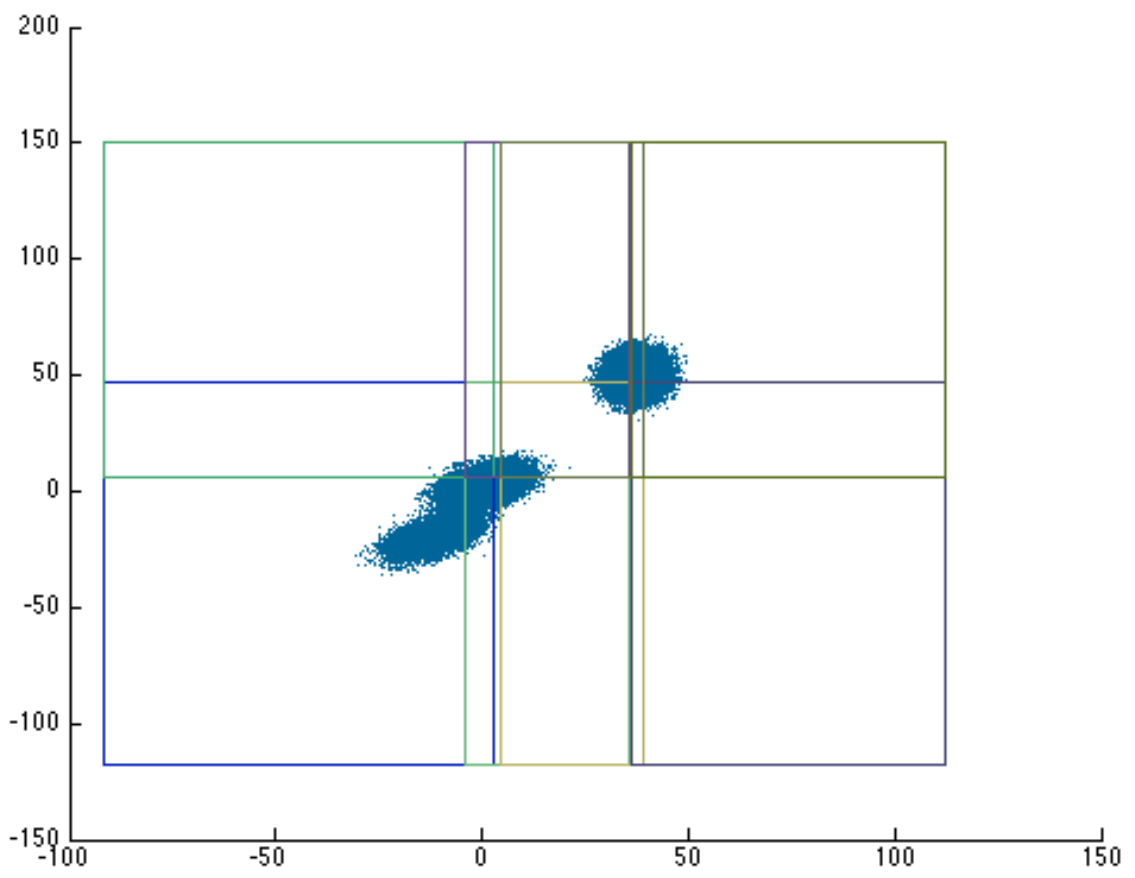

**Fig. 6** Automated Estimate of a two dimensional GMM

The function Quantiles$(x(j), n)$ outputs $n$ quantiles of $x$ in dimension $n$ so that the state space of dimension $j$ is divided in parts where there is equal probability of landing in each part. The quantiles are given by the empirical cumulative distribution.

The function MergeCover takes the individual covers for each dimension and merges them to a cover of the full state space. In doing so it asserts, sequentially, that there is an overlap so that the cover becomes a linked cover.

The variance of $\hat{C}$ depends on the quantile estimates from $\xi^{\text{Pilot}}$. For instance if $\xi^{\text{Pilot}}$ is sampled using Metropolis–Hastings the variance can be found in Nummelin (2002) and decreases with a rate of $\sqrt{M_{\text{Pilot}}}$. In Figure 7 there are $10^5$ covers estimated on a same gamma distribution. A typical examples of a simulation using one of the estimated covers is seen in Figure 5. In Figure 6 a cover for a two dimensional Gaussian mixture model is estimated. As seen, the automated cover generation divides the statespace in the modes of the distribution.

In Figure 8 a five dimensional Gaussian mixture model with several components is simulated. The cover comprising 16 parts is first created with a pilot simulation. Thereafter is rejection sampling used on each of the subcovers. In the Figure 8 the final result from the rejection sampling is displayed. Clearly, most of the samples are generated by few of the workers. On the other hand this means that there is a high resolution in the tail of the distribution.

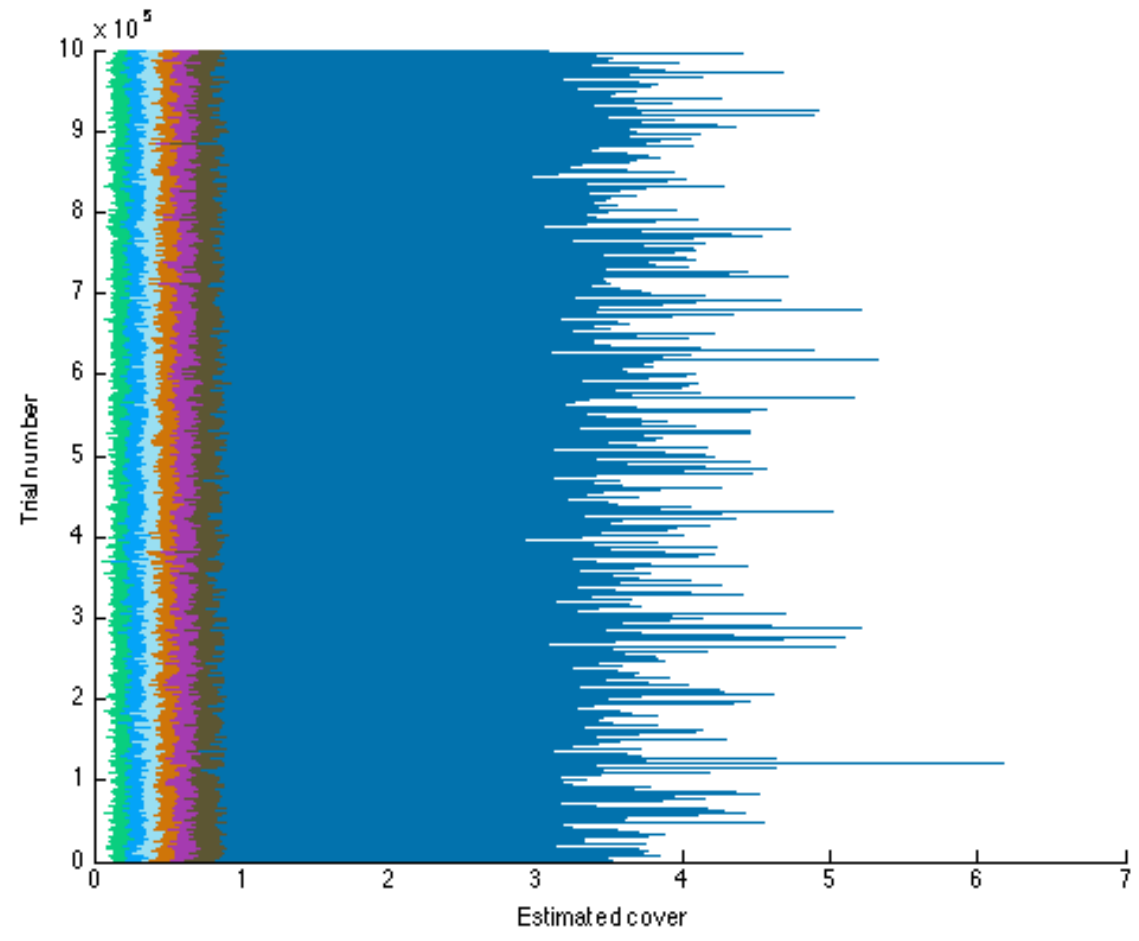

**Fig. 7** Repeated estimation of covers for the gamma distribution

### *2.3.1 Guidelines for implementation*

Given a few workers, each equipped with a GPU, the way to implement DC-sampling would be to first make sure that the GPUs are fully used on each computer and then distribute the computing, letting each worker simulate from a different parts of the sample space. This means that for instance in Henriksen et al (2012) the particle filter parallelization would remain intact while the PMMH sampler would be affected.

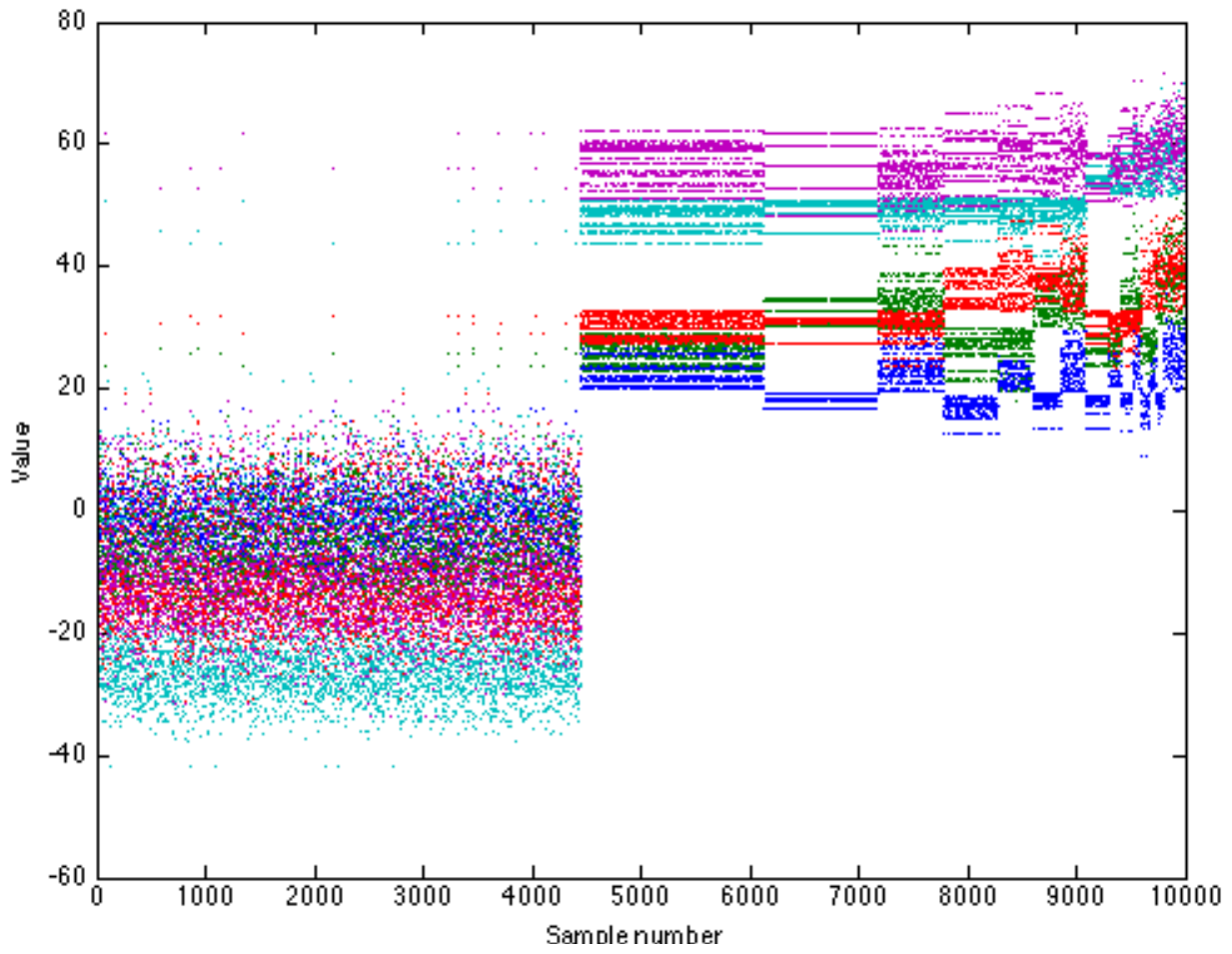

**Fig. 8** Five dimensional GMM with 16 part cover. Each of the 16 parts can be distinguished. Note that this is not a plot of a Markov chain but 16 independent samples. The length of each part represents its relative importance weight so approximately 45% of the samples are taken from the first subcover.

## 3 Applications and results

In this section the two main examples are presented. The first example was briefly discussed in Section 2. The second example utilizes particle Markov chain Monte Carlo. A third example in Section 3.3 reproduces a numerical demonstration from Hastings original paper.

In the first example, the discrete state space, the purpose of the demonstration is to illustrate the advantages of decomposing a state space when it is difficult to move from one part to another. The second case is the PMMH sampler applied to calibration of the volatility model from the previous section. It utilizes the properties of the algorithm to gain more samples in less time by running two chains in parallel. It should be noted that some of the simulations are not run in actual batch mode but instead on a single computer in sequence. The execution time is computed as the average time of the batches.

The number of samples, $N$, will be stochastic and depend on the selection of the cover for the Decomposition sampler. Therefore, after running the Decomposition sampler, the other methods were configured to produce equally many samples. An exception is the following section with the discrete model where the total valuation norm is evaluated against number of samples. The time and total variation results in Table 1 are reported in order of magnitude.

The parallelization was implemented by sending each subproblem to a worker. The parallelization was emulated, meaning that the subproblems were actually executed on a single worker but afterwards treated as if they came from separate workers. The timing was done separately for each emulated parallelism. The topic is further discussed in Youssfi et al (2010).

The emulation does not consider overhead cost but that cost will be small for DC-sampling. For example in the calibration of the volatility model it takes longer to do one single iteration of the MCMC-algorithm than to do the merging process in the end. During the simulation there is no communication between chains. Therefore the overhead cost is ignored througout the paper.

## 3.1 Discrete Sample Space

Consider a Markov chain moving on the discrete space according to Figure 3. The transition probabilities are given by

$$P = \begin{bmatrix} \frac{1}{3} & \frac{1}{3} & \frac{1}{3} & 0 & 0 & 0 & 0 \\ \frac{1}{3} & \frac{1}{3} & \frac{1}{3} & 0 & 0 & 0 & 0 \\ \frac{1-a}{3} & \frac{1-a}{3} & \frac{1-a}{3} & a & 0 & 0 & 0 \\ 0 & 0 & \frac{1-a}{2} & \frac{1-a}{2} & a & 0 & 0 \\ 0 & 0 & 0 & a & \frac{1-a}{3} & \frac{1-a}{3} & \frac{1-a}{3} \\ 0 & 0 & 0 & 0 & \frac{1}{3} & \frac{1}{3} & \frac{1}{3} \\ 0 & 0 & 0 & 0 & a & a & 1-2a \end{bmatrix}. \quad (5)$$

In the numerical experiments two different cases are considered. The first has a smaller value of $a$ and the second has an $a$ which is ten times larger and thus a bit more suitable for the standard MH sampler. In both models the probability of moving from state 1 or 2 to state 1,2 or 3 are equal to one third. There is a low probability to move from the third state to state 4. In the nice case it is 3 ‰ and in the bad case it is ten times lower. Once in state 4 there is a low probability of going to the fifth state and a large probability of jumping back.

### *3.1.1 Numerical Results for the Discrete Model*

Consider a Markov chain moving on the discrete space according to Figure 3. The aim is to obtain samples from the chain.

We implement an MH sampler where the kernel of the chain is used as a proposal distribution. Since the chain is reversible, see Häggström (2002), we will have acceptance rate 1 for the MH sampler. In other words, the MH sampler collapses to simply simulating the Markov chain. In the case when we split the state space the rejection of a proposal always leads back to state 4. That is, the only state that can propose illegal samples is the fourth state. Whenever that happens the chain will stay put. Note that the cover was selected with knowledge of the distribution. With a much larger number of states no prior knowledge it would have been more difficult to find a good cover.

We use three different types of samplers: the standard MH sampler, denoted MH-Standard; the Decomposition sampler, denoted MH-DC; and the method where chains are run in parallel by Rosenthal (2000), denoted MH-R. The Rosenthal-sampler runs parallel copies of chains and then, after removing burnin, combines the samples to one. It is therefore very easy to implement.

The two parallel Rosenthal chains are started in separate parts of the cover for a fair comparison to the DC-sampler. Two agents are engaged, so $W = 2$.

In Table 1 we compare the empirical total variation between the actual stationary distribution $\lambda$ and our estimate $\hat{\lambda}$,

$$\|\hat{\lambda} - \lambda\|_{\mathrm{TV}} = \max_j \left| \frac{1}{N} \sum_{k=1}^{N} \mathbb{I}(\xi_k = j) - \lambda_j \right|,$$

where $\{\xi_k\}_{k=1}^N$ is the generated chain. In the column "states" the states visited by the chain are registered. Clearly, the Decomposition sampling, denoted by MH-DC in the table, is superior to the standard MH method. We obtain a better result in less than a second with the parallel version than is possible in an hour with the standard-version. The alternative methods need much more samples (and thus more time) to obtain the same total variation as the DC-sampling. For example MH-R performs worse with $2 \cdot 10^8$ samples than MH-DC does with little over $10^4$.

The number of samples and times are chosen so that the total variations are comparable. But collecting for example $\mathcal{O}(10^n)$ samples would take the same amount of time for each of the methods. Because of the parallelization there will be a scaling factor improving the speed of MH-DC and MH-R.

The results above the line comes from the vicious distribution in Section 3.1 while the results below the line originates from a moderately more pleasant distribution. That is, the results above the line solve a more difficult problem. That $N$ varies is for DC sampling due to the fact that some samples are discarded. For every simulation there were really $2 \cdot 10^{\cdot}$ samples generated. Because of the selection of the cover, there is a mere 10% increase in samples compared to running the standard sampler; despite that, the convergence is *much* faster. Rosenthal's method always generates a full standard chain for every worker.

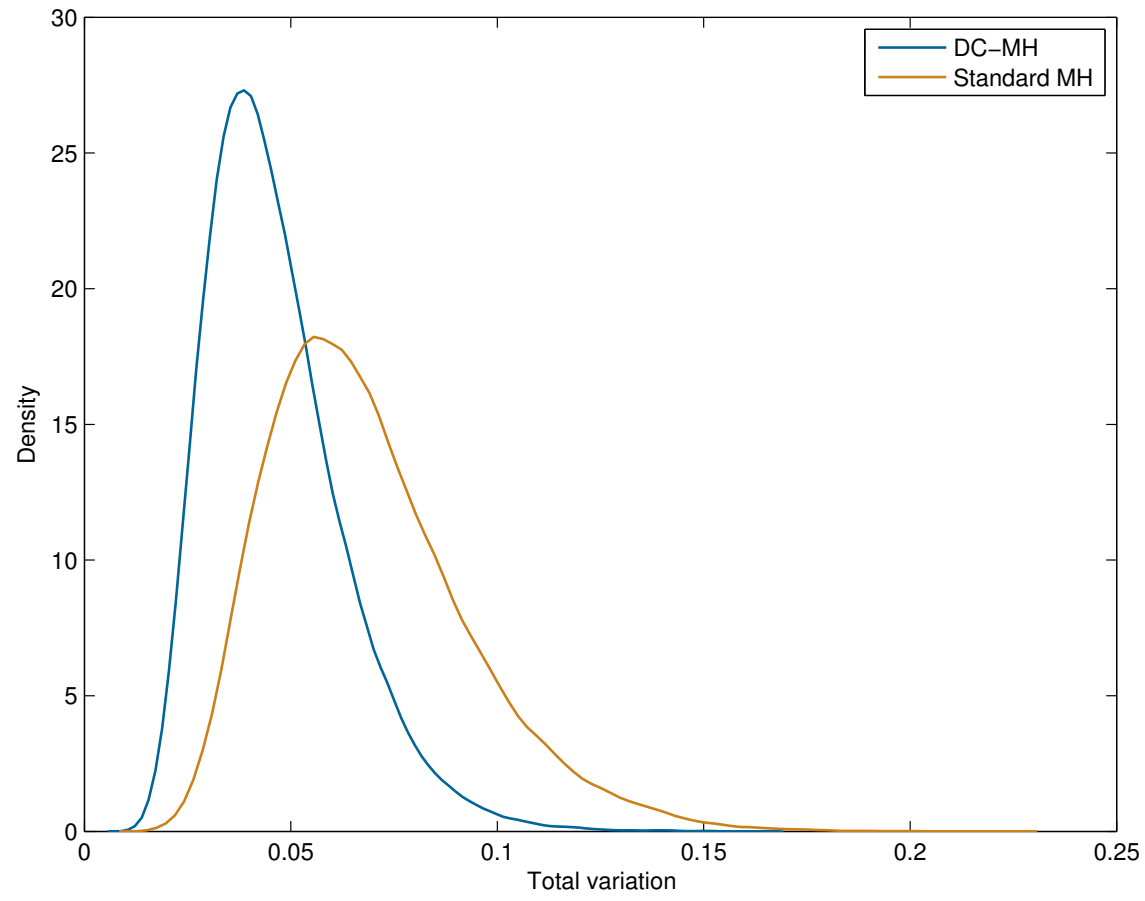

**Fig. 9** Hastings sampled a poisson distribution. Here the distribution was estimated $10^5$ times and an estimate of the total variation is plotted. DC-sampling clearly reduces total varation compared to standard MH in the example.

Running the simulations in the case with the difficult distribution, we find that the standard MH sampler usually does not find its way to the higher numbered states unless the number of iterations is very high.

There are 10% more samples from the parallel chains. The extra samples are the ones gained from the parallelization. The increase in terms of number of samples is quite small, it could have been made larger by splitting in another state than 4; however, the purpose of this experiment was to demonstrate other advantages of the algorithm than the extra samples it provides.

## 3.2 Parallelization of the Particle Marginal MH sampler

The PMMH sampler introduced by Andrieu et al (2010) can solve problems in parameter estimation that were previously unsolved; however, as with other MCMC methods it is not trivial to parallelize. The Decomposition sampling can parallelize any random sampling approach and we choose to demonstrate the algorithm on the PMMH sampler. It will be used to calibrate a stochastic volatility model. The model is for the variation (or the *volatility*) of the price of a financial instrument. This is done in discrete time where we observe the instrument price $S_k$ at time $k$. The volatility model is for the *log returns*, defined as $Y_k = \log(\frac{S_k}{S_{k-1}})$. The model for the log returns is a partially observed hidden

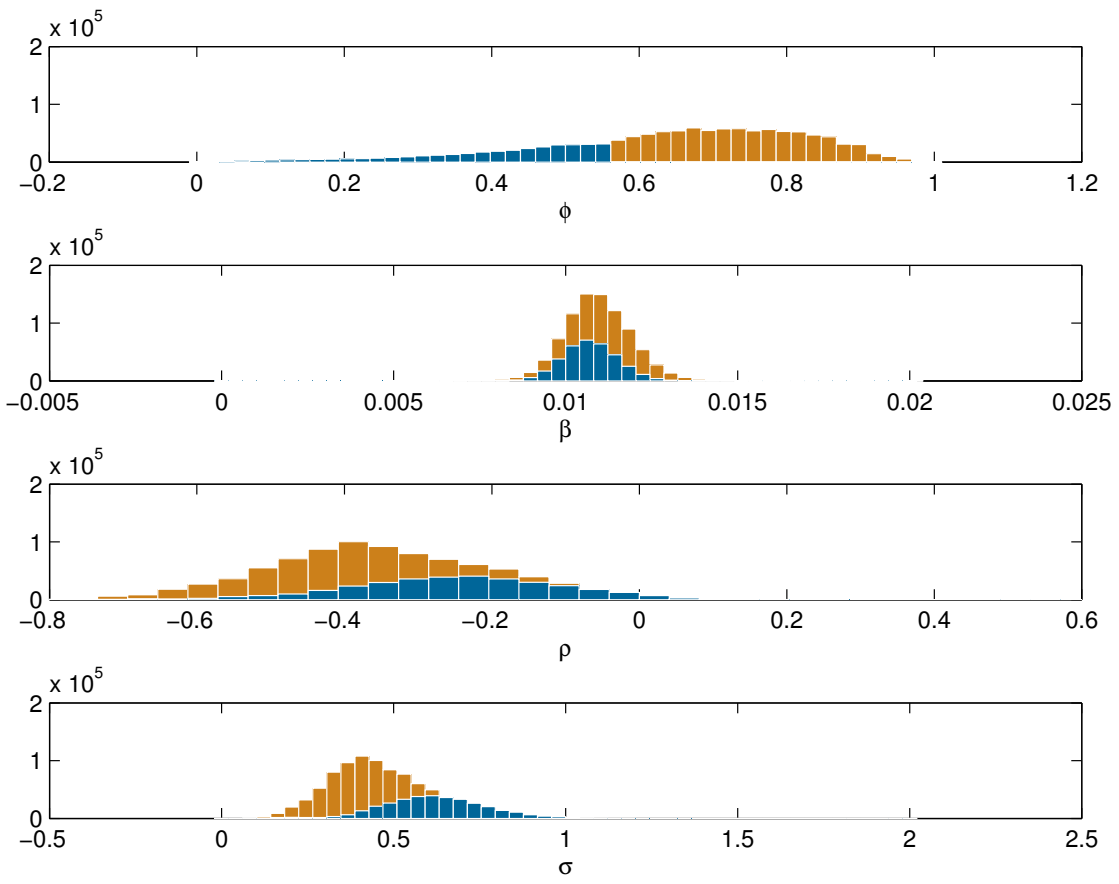

**Fig. 10** PMMH samples for the volatility model from $C_1$ and $C_2$.

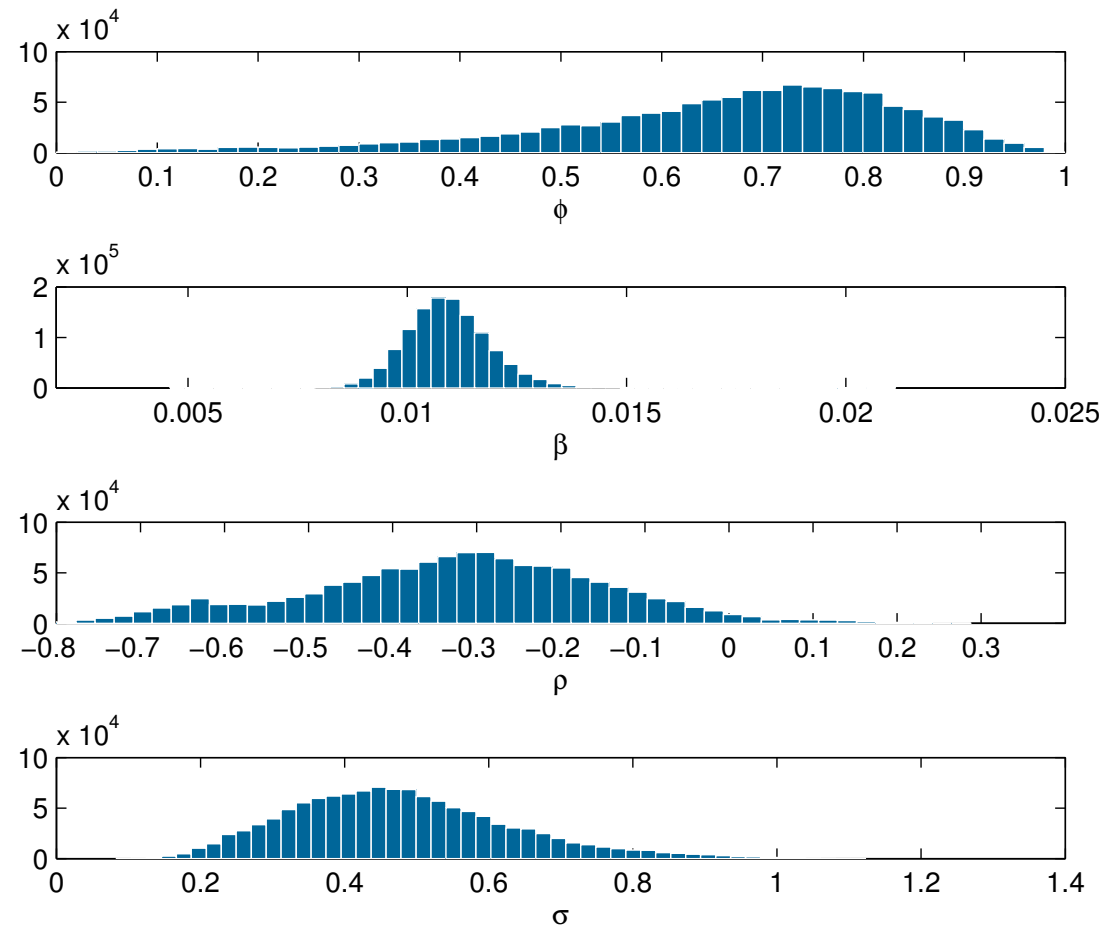

**Fig. 11** Samples from standard PMMH sampling for the volatility model

Markov model

$$\begin{aligned} Y_k &= \beta e^{X_k/2} u_k, \\ X_k &= \phi X_{k-1} + \sigma w_k, \end{aligned} \tag{6}$$

where $Y_k$ are the observed log returns and $X$ is the hidden process driving $Y$.

We fit the model to log returns of the Disney (NYSE:DIS) stock price from the full year of 2007. In figures 10 and 11 we see the resulting samples from the two subsets; they are found in Equation (22) in the appendix. By processing the data through the two algorithms we obtain the final sample in Figure 12. A note for this section is that it is an illustrative example. It is possible to parallelize the particle filter itself as done by Lee et al (2010) which perhaps should be the first measure. The methodology presented in this section could then be used to distribute the computations over several computers.

In Figure 13 the trajectory from the sampler is displayed. The plots in Figure 14 the decaying autocorre-lations indicate that the DC-sampler performs better.

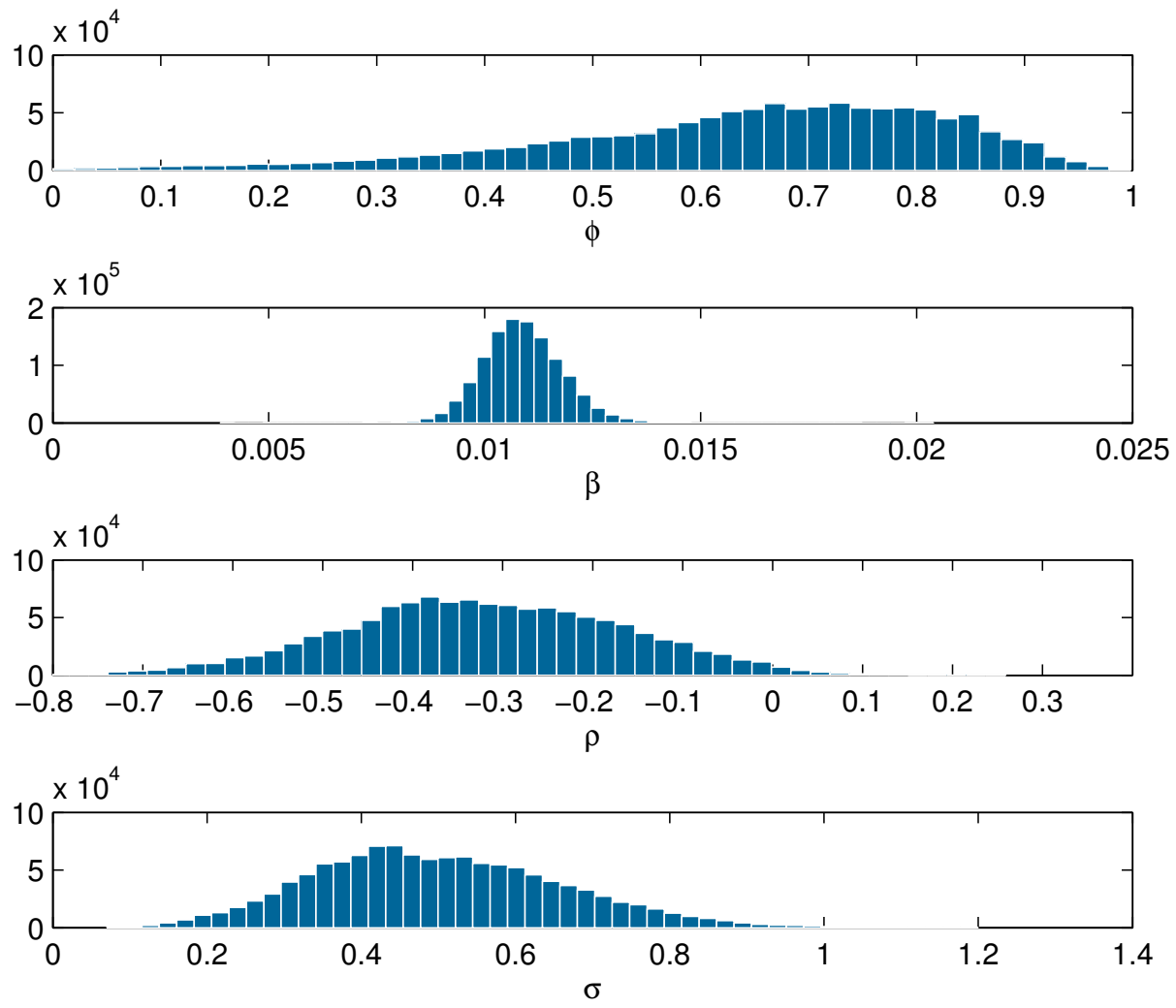

**Fig. 12** Processed PMMH samples from Figure 10

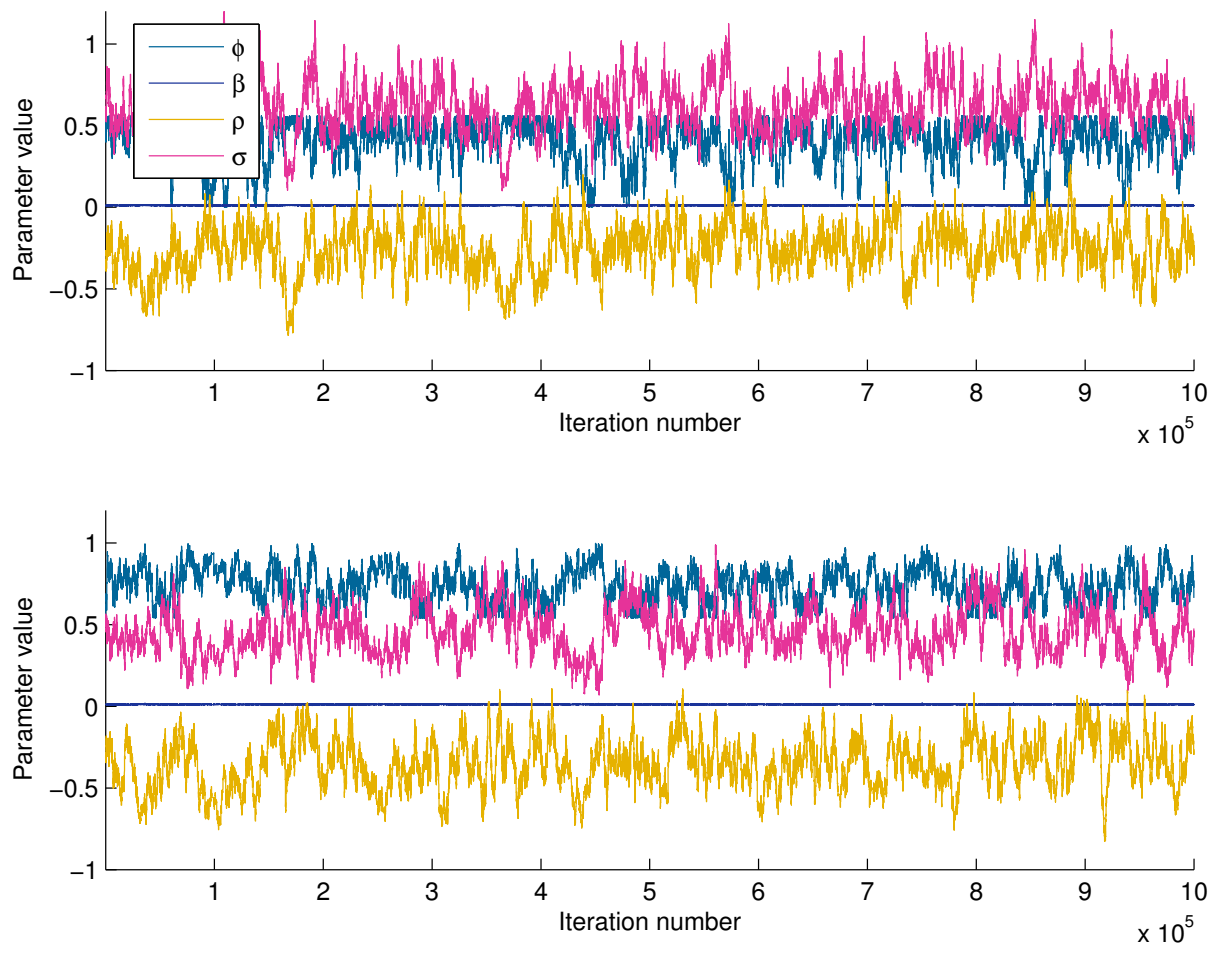

**Fig. 13** Trajectories from PMMH-sampling. The upper plot is the result from $C_1$ and the lower from $C_2$.

### *3.2.1 Calibration of the Stochastic Volatility Model*

We use the split from (22) and compare the result to running a standard MH sampler. The results are presented in Table 2. All the parameters are following a random walk. Initial simulations were run to obtain a covariance matrix for the proposals. The random walk is truncated in the sense that the prior will assign probability zero to any sample outside of the space. This applies to the $\phi$-parameter as well. The Decomposition sampling is run with two workers producing $10^6$ samples each. In the case when the model is calibrated to simulated data roughly 60% of the samples from the first subset are kept resulting in $1.6 \cdot 10^6$ samples in total. As a comparison, the same amount of samples is generated using the standard method. As expected it takes about 1.6 times longer to produce the same amount of samples. In the case with the Disney data, it takes about 1.2 times longer.

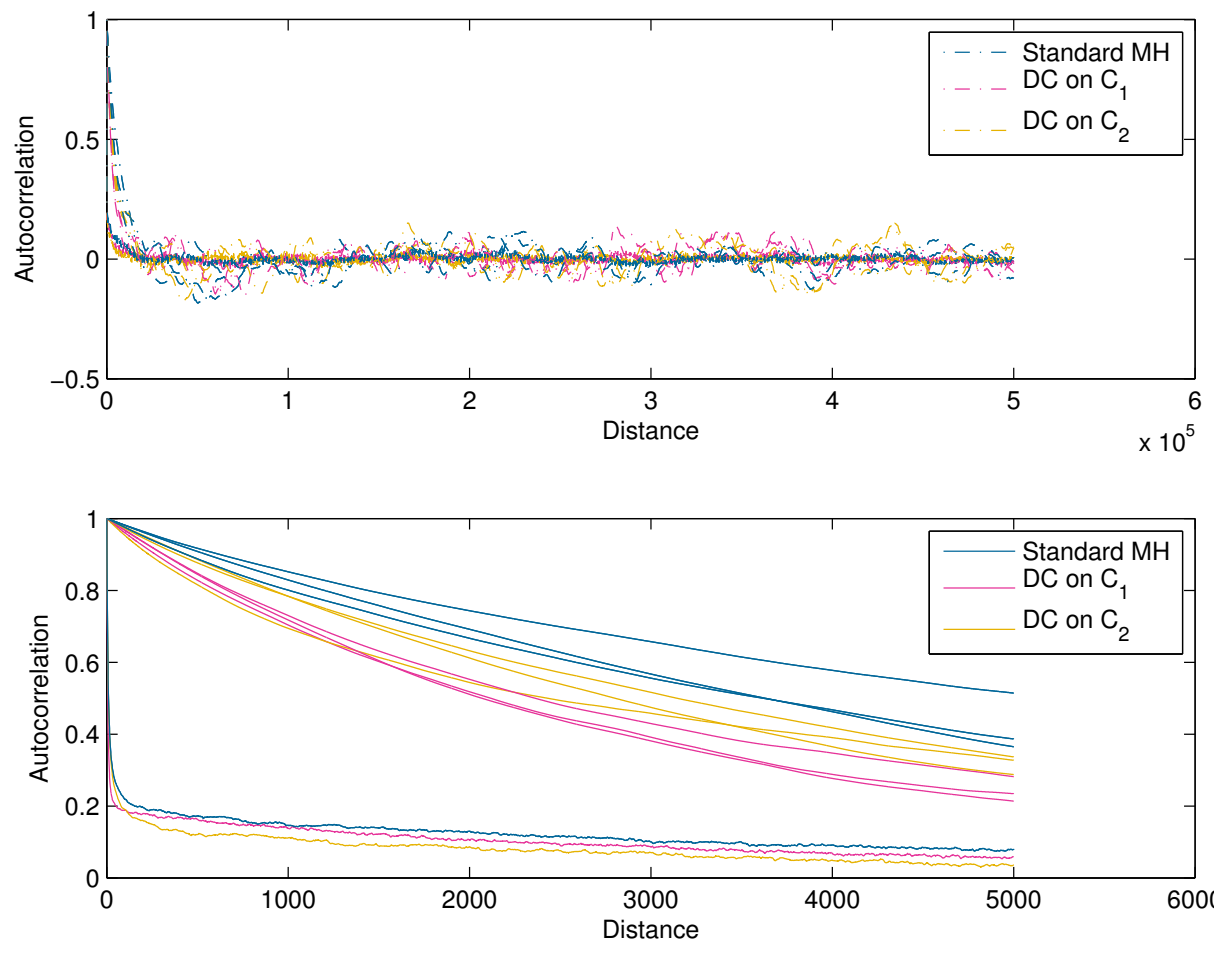

**Fig. 14** Autocorrelation plots for the four parameters from the PMMH-sampling. The upper plot shows the decay of autocorrelation as the distance between two samples in the chain increases. The lower plot is a close-up illuminating that the autocorrelation of the output from the DC-sampler decays faster for all parameters on both parts $C_1$ and $C_2$.

**Table 1** Simulation results for the discrete space models. Simulations below the line have a large value of $a$.

| Method | $N$ | Runtime (s) | $\lVert\hat{\lambda}-\lambda\rVert_{TV}$ | $TV \cdot N$ |
|---|---|---|---|---|
| MH-Standard | $10^8$ | $10^4$ | $10^{-3}$ | $10^5$ |
| MH-Standard | $10^7$ | $10^3$ | $10^{-1}$ | $10^6$ |
| MH-R | $2 \cdot 10^8$ | $10^4$ | $10^{-2}$ | $10^6$ |
| MH-R | $2 \cdot 10^5$ | 10 | $10^{-1}$ | $10^4$ |
| MH-DC | $1.1 \cdot 10^4$ | 1 | $10^{-2}$ | $10^2$ |
| MH-Standard | $10^7$ | $10^3$ | $10^{-2}$ | $10^5$ |
| MH-Standard | $10^6$ | $10^2$ | $10^{-1}$ | $10^5$ |
| MH-R | $2 \cdot 10^8$ | $10^3$ | $10^{-2}$ | $10^6$ |
| MH-R | $2 \cdot 10^6$ | $10^2$ | $10^{-1}$ | $10^5$ |
| MH-DC | $1.1 \cdot 10^7$ | $10^3$ | $10^{-4}$ | $10^3$ |
| MH-DC | $1.1 \cdot 10^3$ | $10^{-1}$ | $10^{-2}$ | $10^1$ |

## 3.3 Hastings Example

In the classic paper by Hastings (1970) an MH sampler is implemented to generate samples from a Poisson distribution. We reproduce this example and compare the performance of the standard MH sampler to that of the Decomposition sampler. The intensity is set to 14 and for each batch 1000 iterations are executed. The state space is split in two parts, where the mode of the distribution is the point of the split. The result is that the total variation is significantly smaller for the Decomposition sampler than for the MH sampler. Repeating $10^5$ such batches yields the estimate of the TV norm in Figure 9.

**Table 2** Simulation results for the volatility model

| Method | Data | N | Runtime (s) |
|---|---|---|---|
| MH-Standard | Simulated | $1.6 \cdot 10^6$ | $1.8 \cdot 10^4$ |
| MH-DC | Simulated | $1.6 \cdot 10^6$ | $1.1 \cdot 10^4$ |
| MH-Standard | NYSE:DIS | $1.2 \cdot 10^6$ | $1.2 \cdot 10^4$ |
| MH-DC | NYSE:DIS | $1.2 \cdot 10^6$ | $1.1 \cdot 10^4$ |

In this case it was easy to design a proposal distribution.

The theoretical results in Appendix A are well illustrated in the reduction of total variation. The lower total variance is due to the reduction of the state space which let the Decomposition sampler be more careful in its exploration of the limited parts of the state space.

### 3.4 Software Package and implementation details

A software package is provided as a digital attachment. It consists of a toolbox for automated cover selection and DC-sampling. The only input needed is the function to sample on the subcovers. The examples from the paper are also included so further details about the implementations are found there.

### 3.5 Discussion and Developments

The algorithm has at least two interesting uses: parallelization and custom designed samplers. The parallelization yields, if the cover $C$ is carefully selected, more samples. Sometimes, as is very much the case with Hastings example, the gain is not in the number of samples we get but in the polytropic exploration of the state space. The automated cover selection has a design parameter, $\delta$. Further investigations of theoretical results could give better methods for selecting the parameter.

A Rao-Blackwellization, see Casella and Robert (1996), of Algorithm 1 would allow a more frugal use of the plentiful samples. The discrete spaced example from section 3.1 is similar to the Wang–Landau example by Fort et al (2014). A performance comparison between Decomposition sampling and other methods mentioned in the introduction would be interesting.

Problems and limitations of DC-sampling lies mainly in the selection of the cover which affects the performance. The simple example with the gamma distribution demonstrated that cover selection by pilot simulation is a rather robust process. However as the dimensionality increases it is likely that the robustness decreases. For most practical applications we therefore recommend using as few dimensions as possible for the cover split. For instance, dividing the first 3 dimensions in 4 parts would give 81 subcovers.

Future work should focus on developing the automated cover selection. Further developments of the sampling algorithms would accord greater freedom for the overlaps of the cover. That could stabilize and simplify the coverselection. Several other improvements can be made. The cover could be updated on the go, reusing old observations. For cases when the likelihood can be evaluated without simulation a cover could be found by running an optimization algorithm to find local minma and maxima.

Given the results on selected problems we consider DC-sampling an interesting addition to the Monte Carlo toolbox, in particular for specialized applications.

**Acknowledgements** The authors are thankful to Jimmy Olsson for his guidance in the calibration of the stochastic volatility model and to Tobias Rydén for his comments and suggestions.

# A Theoretical Results

In this part of the appendix theory supporting the validity of the algorithms is given. The following proposition relates to standard results for sampling random numbers. It establishes convergence in total variation which means that as the number of samples increases the distance between our estimated distribution and the actual distribution will vanish. The second part of the next proposition depend on the concept of *irreducibility*. Let $\{\xi_k\}_{k\geq 1}$ be a Markov chain. If for each set $A \in \mathcal{B}(\mathsf{S})$ with $\pi(A) > 0$ there exists an $n$ such that $\mathbb{P}(\xi_n \in A) > 0$ for every starting value of the chain, we say that the chain is $\pi$-irreducible.

**Proposition 1** *Let $\eta$ be a sample from Algorithm 1.*

1. *If $\int_\cdot q_k^{C_j}(x)\,dx = \pi(\cdot \mid C_j)$ for every $j$ and for every $k$, then $\eta$ is a sample from $\mathbb{P}$.*
2. *Let each $\xi^{C_j}$ be $\pi(\cdot \mid C_j)$-irreducible, let $\zeta$ be a MH chain of length $M$ from the full space $\mathsf{S}$, and let $\mu$ be a probability distribution. Then given any $\zeta$ it is possible to construct an $\eta$ such that the total variation norm of the distribution of the $M$'th draw of $\eta$, denoted $\pi_\eta^M(\cdot) = \int \mathbb{P}(\eta_M \in \cdot \mid x)\mu(dx)$, with respect to $\pi$ satisfies the following inequality for $\pi_\zeta$*

$$\|\pi_\eta^M - \pi\|_{\mathrm{TV}} \leq \|\pi_\zeta^M - \pi\|_{\mathrm{TV}} \sum_{j=1}^{W} \pi(C_j), \tag{7}$$

*this expression tends to zero as $M$ tends to infinity.*
*Let $\pi^M(C_j)$ be an estimate such that $\|\pi^M(C_j) - \pi(C_j)\|_{\mathrm{TV}}$ tends to zero as $M$ tends to infinity for every $j$. Use the estimate $\pi^M(C_j)$ for $\mathbb{P}(X \in C_j)$ in Algorithm 1. Then the distribution $\pi_\eta^M$ of the $M$'th draw of $\eta$ converges in total variation to that of $\pi$,*

$$\left\|\pi_\eta^M - \pi\right\|_{\mathrm{TV}} \xrightarrow[M\to\infty]{} 0. \tag{8}$$

By Robert and Casella (2004) a chain generated from the MH algorithm is $\pi$-irreducible under some mild conditions on the transition kernel. Thus, if each $\xi^{C_j}$ is an MH chain, then (7) and (8) hold.

Let $\pi^M = \{\pi^M(C_1), \ldots, \pi^M(C_W)\}$ be a sample from Algorithm 3. Define

$$\pi^M(C_j \setminus \Delta_{1:j-1}) = \pi^M(C_j)\frac{1}{M}\sum_{k=1}^{M} \mathbb{I}(\xi_k^{C_j} \notin \Delta_{1:j-1}).$$

The convergence results for Algorithm 3 differ from that of the previous proposition. Here we consider the mean and the variance of the estimates. That the variance of $\pi^M(C_j)$ tends to zero as $M$ tends to infinity is demonstrated in Figure 4.

**Proposition 2** *Almost surely, for each $j$,*

$$\pi^M(C_j) \xrightarrow[M\to\infty]{} \pi(C_j), \tag{9}$$

*and*

$$\pi^M(C_j \setminus \Delta_{1:j-1}) \xrightarrow[M\to\infty]{} \pi(C_j \setminus \Delta_{1:j-1}). \tag{10}$$

*Furthermore, if for every $j$ $\xi^{C_j}$ is an MH chain which is $\pi(\cdot \mid C_j)$-irreducible, then (9) and (10) still hold. Assuming that the algorithm does not fail then the variances of $\pi^M(C_j)$ and $\pi^M(C_j \setminus \Delta_{1:j-1})$ tend to zero as $M$ tends to infinity:*

$$\mathbb{V}[\pi^M(C_j)] \xrightarrow[M\to\infty]{} 0.$$

The expression for the variance is recursive and is given in Appendix B. In Figure 4 the gamma distribution is used once again to create a cover with three parts. The pilot simulation giving rise to the cover consists of 100 samples. The probabilities are estimated with an increasing number of samples giving reduced variance as a result.

Proposition 2 implies that we can get all the required probabilities by looking at the overlaps. A slightly more general version can be obtained. If each subsample $\xi^{C_j}$ has a unique $M^{C_j}$, then the same result is obtained if we use the mean value in the update step (4) of $\hat{\pi}^M(C_j)$ above by letting the smallest such $M^{C_j}$ tend to infinity. The update turns into

$$\hat{\pi}^M(C_j) = \hat{\pi}^M(C_{j-1})\frac{M^{C_j}\sum_{k=1}^{M^{C_{j-1}}} \mathbb{I}(\xi_k^{C_{j-1}} \in \Delta_{j-1})}{M^{C_{j-1}}\sum_{k=1}^{M^{C_j}} \mathbb{I}(\xi_k^{C_j} \in \Delta_{j-1})}.$$

Finally, we give a proposition saying that the previous Algorithm can be used in combination with Algorithm 2.

**Proposition 3** *Let $\pi^M(C_j \setminus \Delta_{1:j-1})$ be an estimate of $\pi(C_j \setminus \Delta_{1:j-1})$ from Algorithm 3 and let $\eta$ be a sample from Algorithm 2 of size $M$, using that estimate. Further assume that $h \in L^1[\pi(\cdot \mid C_j)]$ for every $j$; then as $M$ tends to infinity,*

$$\frac{1}{M}\sum_{k=1}^{M} \eta_k \longrightarrow \mathbb{E}[h(X)], \tag{11}$$

*almost surely. Furthermore if $\xi^{C_j}$ is an MH chain which is $\pi(\cdot \mid C_j)$-irreducible, then (11) still holds.*

If the true probabilities $\pi(C_j \setminus \Delta_{j-1})$ are known the proposition still holds but Algorithm 3 is redundant.

# B Proofs

## B.1 Proof of Proposition 1

*i)* We will examine an arbitrarily selected $k$. First, note that in the downsampling step we multiply the acceptance probability with the indicator function $\mathbb{I}(\xi_k^{C_j} \notin \Delta_{1:j-1})$; this is the same thing as forcing our draw to be conditionally from $C_j \setminus \Delta_{1:j-1}$. In other words, the probability of an event $A$ is evaluated conditioned that it does not belong to the intersection $\Delta_{1:j-1}$. Thus

$$\mathbb{P}(\eta_k \in A) = \sum_{j=1}^{W} \mathbb{P}(\xi^{C_j} \in A \setminus \Delta_{1:j-1})\mathbb{P}(X \in C_j),$$

where we use that the probability that a distinct $\eta_k$ comes from a certain $C_j$ is precisely $\mathbb{P}(X \in C_j)$. All the $\xi^{C_j}$s are drawn from the conditional distributions $X \mid X \in C_j$ so

$$\mathbb{P}(\eta_k \in A) = \sum_{j=1}^{W} \mathbb{P}(X \in A \setminus \Delta_{1:j-1} \mid X \in C_j)\mathbb{P}(X \in C_j).$$

Since this coincides with the expression in (2) we end up with $\mathbb{P}(\eta_k \in A) = \mathbb{P}(X \in A)$ for an arbitrarily selected $k$ and the property is established.

*ii)* Recall that $\pi(A) = \mathbb{P}(X \in A)$. We want to show

(a) $\pi_\eta^M \xrightarrow{\mathrm{TV}} \pi$

(b) $\|\pi_\eta^M - \pi\|_{\mathrm{TV}} \leq \|\pi_\zeta^M - \pi\|_{\mathrm{TV}} \sum_{j=1}^{W} \mathbb{P}(X \in C_j)$,

where

$$\pi_\eta^M(\cdot) = \int \mathbb{P}(\eta_M \in \cdot \mid x)\mu(dx),$$

$x$ is a vector of initial values for $\xi$ and $\mu$ is a probability distribution. The proposition states that it is possible to construct a sequence such that the inequality holds. We prove the statements in reversed order.

(b) Let $\pi(C_j)$ denote the probability $\mathbb{P}(X \in C_j)$, and let $x_j$ be the initial value of each chain $\xi^{C_j}$, further let $M_j$ denote the current time of each chain when $\eta$ is at time $M$. Consider

$$\begin{aligned}
&\left\|\pi_\eta^M - \pi\right\|_{\mathrm{TV}} = \sup_A |\pi_\eta^M(A) - \pi(A)| \\
&= \sup_A \left| \Big( \sum_{j=1}^{W} \mathbb{P}(\xi_{M_j}^{C_j} \in A \setminus \Delta_{1:j-1} \mid x_j)\pi(C_j) \Big) - \pi(A) \right| \\
&= \sup_A \left| \sum_{j=1}^{W} \Big\{ \mathbb{P}(\xi_{M_j}^{C_j} \in A \setminus \Delta_{1:j-1} \mid x_j) \right. \\
&\quad \left. - \pi(A \setminus \Delta_{1:j-1} \mid X \in C_j) \Big\} \pi(C_j) \right| \\
&\le \sup_A \max_j \left| \Big\{ \mathbb{P}(\xi_{M_j}^{C_j} \in A \setminus \Delta_{1:j-1} \mid x_j) \right. \\
&\quad \left. - \pi(A \setminus \Delta_{1:j-1} \mid C_j) \Big\} \right| \sum_{s=1}^{W} \pi(C_s) \\
&= \max_j \|\pi_{\xi_{C_j \setminus \Delta_{1:j-1}}}^{M_j} - \pi(\cdot \mid C_j \setminus \Delta_{1:j-1})\|_{\mathrm{TV}} \sum_{s=1}^{W} \pi(C_s).
\end{aligned} \tag{12}$$

We see that a sufficient, but not necessary, condition for the inequality to hold is the following assumption.

**Assumption 1** *For $\xi$ and $\zeta$ the inequality*

$$\max_j \|\pi_{\xi_{C_j \setminus \Delta_{1:j-1}}}^{M_j} - \pi(\cdot \mid C_j \setminus \Delta_{1:j-1})\|_{\mathrm{TV}} \le \|\pi_\zeta^{M_j} - \pi\|_{\mathrm{TV}}$$

*holds.*

With that assumption, we can use the final line of (12) to establish the inequality as

$$\left\|\pi_\eta^M - \pi\right\|_{\mathrm{TV}} \le \|\pi_\zeta^{M_j} - \pi\|_{\mathrm{TV}} \sum_{s=1}^{W} \pi(C_s).$$

It is left to show that it is possible to satisfy the assumption for any $\zeta$. The assumption seems strong, but it can be interpreted as "Assume that we do not make a deliberately bad choice of the proposal density". To demonstrate that the assumption always can be fulfilled, let the proposal density of $\xi_{C_j \setminus \Delta_{1:j-1}}$ be the same as that of $\zeta$. That is, samples for $\xi_{C_j \setminus \Delta_{1:j-1}}$ are generated in the same way as for $\zeta$ but only samples in $C_j \setminus \Delta_{1:j-1}$ are included. Then the assumption must be true since

$$\begin{aligned}
&\|\pi_{\xi_{C_j \setminus \Delta_{1:j-1}}}^{M_j} - \pi_{C_j \setminus \Delta_{1:j-1}}\|_{\mathrm{TV}} \\
&= \|\pi_{\zeta_{C_j \setminus \Delta_{1:j-1}}}^{M_j} - \pi_{C_j \setminus \Delta_{1:j-1}}\|_{\mathrm{TV}} \le \|\pi_\zeta^{M_j} - \pi\|_{\mathrm{TV}}.
\end{aligned}$$

Thus, (b) is established.

(a) To prove (a), note that $M_j \ge \frac{M}{W}$ with equality only when we accept every proposal. Since $W$ is fixed, we realize that $M_j$ tends to infinity as $M$ tends to infinity. By Theorem 7.4 in Robert and Casella (2004) we know that if $\xi_{C_j}$ is a $\pi(\cdot \mid C_j)$-irreducible MH Markov Chain, then, for every initial distribution of $x_j$, the norm

$$\|\pi_{\xi_{C_j}}^M - \pi(\cdot \mid C_j)\|_{\mathrm{TV}} \tag{13}$$

tends to zero as $M$ tends to infinity. If scaled with $W$, an upper bound for $\left\|\pi_\eta^M - \pi\right\|_{\mathrm{TV}}$ is seen in Equation (12) since $W \le \sum_{s=1}^{W} \pi(C_s)$ and

$$\|\pi_{\xi_{C_j \setminus \Delta_{1:j-1}}}^{M_j} - \pi(\cdot \mid C_j \setminus \Delta_{1:j-1})\|_{\mathrm{TV}} \le \|\pi_{\xi_{C_j}}^M - \pi(\cdot \mid C_j)\|_{\mathrm{TV}}.$$

The claim follows: the norm $\left\|\pi_\eta^M - \pi\right\|_{\mathrm{TV}}$ tends to zero as $M$ tends to infinity.

In the last part of the proposition $\pi(C_j)$ is assumed to be replaced with the estimate $\pi^M(C_j)$. Define $\varepsilon_j^{(M)} = \pi^M(C_j) - \pi(C_j)$, then $\pi^M(C_j) = \pi(C_j) + \varepsilon_j^{(M)}$. The following is assumed in the proposition.

**Assumption 2** *The total variation distance $\|\varepsilon_j^{(M)}\|_{\mathrm{TV}}$ tends to zero as $M$ tends to infinity.*

Modify the expression from (12) to include

$$\begin{aligned}
&\mathbb{P}(\xi_{M_j}^{C_j} \in A \setminus \Delta_{1:j-1} \mid x_j)\pi^M(C_j) \\
=&\mathbb{P}(\xi_{M_j}^{C_j} \in A \setminus \Delta_{1:j-1} \mid x_j)(\pi(C_j) + \varepsilon_j^{(M)}).
\end{aligned}$$

This gives

$$\begin{aligned}
&\left\|\pi_\eta^M - \pi\right\|_{\mathrm{TV}} \le \sup_A \max_j \left| \Big\{ \mathbb{P}(\xi_{M_j}^{C_j} \in A \setminus \Delta_{1:j-1} \mid x_j) \right. \\
&\quad \left. - \mathbb{P}(X \in A \setminus \Delta_{1:j-1} \mid X \in C_j) \Big\} \right| \sum_{s=1}^{W} \pi(C_s) \\
&\quad + \|\varepsilon_j^{(M)}\|_{\mathrm{TV}} \sum_{s=1}^{W} \sup_A \mathbb{P}(\xi_{k_s}^{C_s} \in A \setminus \Delta_{s-1} \mid x_s) \\
&\le W \Big( \max_j \|\pi_{\xi_{C_j \setminus \Delta_{1:j-1}}}^{M_j} - \pi(\cdot \mid C_j \setminus \Delta_{1:j-1})\|_{\mathrm{TV}} \\
&\quad + \|\varepsilon_j^{(M)}\|_{\mathrm{TV}} \Big).
\end{aligned}$$

By (13) the norm $\|\pi_{\xi_{C_j \setminus \Delta_{1:j-1}}}^{M_j} - \pi(\cdot \mid C_j \setminus \Delta_{1:j-1})\|_{\mathrm{TV}}$ tends to zero as $M$ tends to infinity and by assumption $\|\varepsilon_j^{(M)}\|_{\mathrm{TV}}$ tends to zero as $M$ tends to infinity. Concluding, under Assumption 2, the estimate $\pi_\eta^M$ converges in total variation to $\pi$.

## B.2 Proof of Proposition 2

This proof relies on the independence of the subsamples.

*Proof* Note that the strong law of large numbers, or when $\xi^{C_j}$ is a $\pi(\cdot \mid C_j)$-irreducible MH chain Theorem 7.4 in Robert and Casella (2004), implies that $(M)^{-1} \sum_{k=1}^{M} \mathbb{I}(\xi_k^{C_j} \in \Delta_j) \xrightarrow[M\to\infty]{} \mathbb{P}(X \in \Delta_j \mid X \in C_j)$, almost surely. Thus, by continuity,

$$\frac{\sum_{k=1}^{M} \mathbb{I}(\xi_k^{C_{j-1}} \in \Delta_j)}{\sum_{k=1}^{M} \mathbb{I}(\xi_k^{C_j} \in \Delta_j)} \xrightarrow[M\to\infty]{} \frac{\pi(\Delta_j \mid C_{j-1})}{\pi(\Delta_j \mid C_j)}$$

almost surely. Put $\hat{\pi}^M(C_1)=1$ and update the other quantities $\hat{\pi}^M(C_j)$. The estimated distribution $\hat{\pi}^M(C_j)$ is now proportional to the true distribution $\pi$, which is such that $\pi(C_j)=\mathbb{P}(X\in C_j)$. The sets in $C$ are pairwise intersected, these intersections are counted twice, therefore

$$\sum_{j=1}^{W}\pi(C_j)=1+\sum_{j=1}^{W}\pi(\Delta_{1:j-1})$$
$$=1+\sum_{j=1}^{W}\pi(\Delta_{1:j-1}\mid C_j)\pi(C_j).$$

In other words, the sum $\sum_{j=1}^{W}\pi(C_j)$ can be adjusted to match a probability distribution, meaning that it sums to 1,

$$\sum_{j=1}^{W}\pi(C_j)\{1-\pi(\Delta_{1:j-1}\mid C_j)\}=1.$$

We exploit this relationship to obtain the normalizing constant. This proves the convergence in (9). The second convergence follows from taking the limit of the left-hand side of (10). It is given by

$$\begin{aligned}\mathbb{P}(X\in C_j)\mathbb{P}(X\notin\Delta_{1:j-1}\mid X\in C_j)&=\mathbb{P}(X\in C_j\cap X\notin\Delta_{1:j-1})\\&=\mathbb{P}(X\in C_j\setminus\Delta_{1:j-1}),\end{aligned}$$

which coincides with the right-hand side of (10).

## B.3 Convergence of the Variance in Proposition 2

To prove the statement we start with a few lemmas establishing properties for binomial random variables. Then, in Section B.3.1, the lemmas are used to establish the result.

**Lemma 1 (Two Recursions)** *Let $t_j=t_{j-1}r_j$ and $t_1=r_1$. Then*

$$t_j=\prod_{s=1}^{j-1}r_s.$$

*Let $T_j=T_{j-1}R_j+c_j$ and $T_1=R_1$, then*

$$T_j=\prod_{s=1}^{j}R_s+\sum_{s=0}^{j-3}\left\{c_{j-s-1}\prod_{u=j-s}^{j}R_u\right\}+c_j.$$

*Proof* Note that $t_2=t_1r_2=r_1r_2$. Also note that if $t_j=\prod_{s=1}^{j-1}r_s$ then

$$t_{j+1}=r_{j+1}\prod_{s=1}^{j-1}r_s=\prod_{s=1}^{j}r_s.$$

This proves, by induction, that the first recursion holds. In the same manner as for the first recursion, note that

$$T_2=T_1R_2+c_2=R_1R_2+c_2.$$

If $T_j=\prod_{s=1}^{j}R_s+\sum_{s=0}^{j-3}\left\{c_{j-s-1}\prod_{u=j-s}^{j}R_u\right\}+c_j$ then

$$\begin{aligned}T_{j+1}&=R_{j+1}\Big(\prod_{s=1}^{j}R_s+\sum_{s=0}^{j-3}\Big\{c_{j-s-1}\prod_{u=j-s}^{j}R_u\Big\}+c_j\Big)+c_{j+1}\\&=\prod_{s=1}^{j+1}R_s+\sum_{s=0}^{j-3}\Big\{c_{j-s-1}\prod_{u=j-s}^{j+1}R_u\Big\}+R_{j+1}c_j+c_{j+1}\\&=\prod_{s=1}^{j+1}R_s+\sum_{s=0}^{j-3}\Big\{c_{j+1-s-1}\prod_{u=j+1-s}^{j+1}R_u\Big\}+c_{j+1},\end{aligned}\tag{14}$$

which concludes the lemma.

Lemmas 2 and 3 are considered known but the proofs are given for completeness. In the lemmas we discuss properties of a binomial random variable $Z$. We also consider $Z\mid Z>0$, called a *zero truncated binomial variable*

**Lemma 2 (Zero Truncated Binomial Variables)** *Let $Z$ be a binomial variable with parameters $M$ and $\nu$.*

1. *Central moments Let $\mu_k$, and $\gamma_k$ denote the k'th central moment of $Z$ and $Z\mid Z>0$ respectively. Then*

$$\begin{aligned}(1-(1-\nu)^M)(-1)^k\mathcal{O}(\frac{\gamma_k}{\gamma^{k+1}})&=(-1)^k\mathcal{O}(\frac{\mu_k}{\mu^{k+1}})\\&=(-1)^k\mathcal{O}(\frac{M^{\lfloor k/2\rfloor}}{M^{k+1}}).\end{aligned}$$

2. *Expectation of inverse moments Define $\mu=M\nu$. Then*

$$\frac{1}{\mu}\le(1-(1-\nu)^M)\mathbb{E}[Z^{-1}\mid Z>0]=\frac{1}{\mu}+\mathcal{O}(M^{-2}),$$
$$\frac{1}{\mu^2}\le(1-(1-\nu)^M)\mathbb{E}[Z^{-2}\mid Z>0]=\frac{1}{\mu^2}+\mathcal{O}(M^{-3}).$$

*Proof* Consider the recursion

$$\mu_k=M(k-1)\nu(1-\nu)\mu_{k-2}+\nu(1-\nu)\frac{\partial}{\partial\nu}\mu_{k-1}$$

given by Riordan et al (1937). It implies that the order increases only every other moment, or put differently $\mu_k=\mathcal{O}(M^{\lfloor k/2\rfloor})$. For example since $\mu_2$ is of order $\mathcal{O}(M)$ so is $\mu_3$. Consequently

$$\mathcal{O}\Big(\frac{\mu_k}{\mu^{k+1}}\Big)=\mathcal{O}\Big(\frac{M^{\lfloor k/2\rfloor}}{M^{k+1}}\Big).\tag{15}$$

Consider the relation between $\gamma$ and $\mu$. The probability that $Z$ is zero, given as $(1-\nu^M)$, implies that the relation $\gamma_k=\frac{\mu_k-(-\mu)^k(1-\nu)^M}{1-(1-\nu)^M}$ holds. This decides the order of the moments. Observe that

$$\begin{aligned}&(1-(1-\nu)^M))(-1)^k\gamma_k=(-1)^k(\mu_k-(-\mu)^k(1-\nu)^M)\\&=(-1)^k(\mu_k-(-1)^k\mu^k(1-\nu)^M)\\&=((-1)^k\mu_k-(-1)^{2k}\mu^k(1-\nu)^M)\\&=((-1)^k\mu_k-\mu^k(1-\nu)^M)<(-1)^k\mu_k.\end{aligned}\tag{16}$$

Further, since $\gamma=\frac{\mu}{1-(1-\nu)^M}$ we have that $\gamma^{-k}\le\mu^{-k}$ and as a result

$$\mathcal{O}(\gamma^{-k})\le\mathcal{O}(\mu^{-k}).$$

Combining this with (16) gives

$$(1-(1-\nu)^M)(-1)^k \frac{\gamma_k}{\gamma^{k+1}} \le (-1)^k \frac{\mu_k}{\gamma^{k+1}}. \tag{17}$$

By the reasoning above and Equation (15) we get

$$\mathcal{O}\Big(\frac{\mu_k}{\gamma^{k+1}}\Big) \le \mathcal{O}\Big(\frac{M^{\lfloor k/2\rfloor}}{M^{k+1}}\Big).$$

Using this with the inequality in (17) finally gives

$$(1-(1-\nu)^M)(-1)^k \frac{\gamma_k}{\gamma^{k+1}} = (-1)^k \mathcal{O}(M^{-\lfloor k/2\rfloor -1}).$$

This concludes the first part of the lemma.

The two lower bounded inequalities in the second part are obtained by applying Jensen's inequality to the convex functions $\mu^{-1}$ and $\mu^{-2}$.

For the two upper bounded cases, note that by Taylor's theorem $g(Z) = g(\mu) + \sum_{k=1}^{\infty} g^{(k)}(\mu)\frac{(Z-\mu)^k}{k!}$, where $g^{(k)}$ denotes the $k$th derivative. Taking expectation and denoting $\mathbb{E}[(Z-\mu)^k]$ by $\mu_k$ gives

$$\mathbb{E}[g(Z)] = g(\mu) + \sum_{k=2}^{\infty} g^{(k)}(\mu)\frac{\mu_k}{k!}.$$

Define $\gamma = \mathbb{E}[Z \mid Z>0] = (1-(1-\nu)^M)\mu$. In the first case we have $g_1 : x \mapsto x^{-1}$ where $g_1^{(k)} = \frac{(-1)^k k!}{\gamma^{k+1}}$. We see that the truncation is needed for the expectations to be finite. By Taylor's theorem $\mathbb{E}[g_1(Z) \mid Z>0] = \frac{1}{\gamma} + R_1$ where the remainder $R_1$ is given by

$$\sum_{k=2}^{\infty}(-1)^k \frac{k!}{\gamma^{k+1}}\frac{\gamma_k}{k!} = \sum_{k=2}^{\infty}(-1)^k \frac{\gamma_k}{\gamma^{k+1}}.$$

Using the first part of the lemma we find that

$$\begin{aligned}(1-(1-\nu)^M)R_1 &= \sum_{k=2}^{\infty}(-1)^k \mathcal{O}\left(\frac{M^{\lfloor k/2\rfloor}}{M^{k+1}}\right)\\ &= \frac{1}{M}\sum_{k=2}^{\infty}(-1)^k\mathcal{O}(M^{\frac{k-2k}{2}})\\ &= \frac{1}{M}\sum_{k=2}^{\infty}(-1)^k\mathcal{O}(M^{-k/2}) = \mathcal{O}(M^{-2}),\end{aligned}$$

where the series converges by Leibniz criterion. This implies that

$$R_1 = \frac{1}{1-(1-\nu)^M}\mathcal{O}(M^{-2}).$$

In the second case $g_2 : x \mapsto x^{-2}$ and

$$g_2^{(k)}(\gamma) = \frac{(-1)^{k+1}(k+1)!}{\gamma^{k+2}}.$$

Here $\mathbb{E}[g_2(Z)] = \frac{1}{\gamma^2} + \bar{R}_1$. The remainder $\bar{R}_1$ is defined as

$$\sum_{k=2}^{\infty}(-1)^{k+1}\frac{(k+1)!}{\gamma^{k+2}}\frac{\gamma_k}{k!} = \sum_{k=2}^{\infty}(-1)^{k+1}\frac{k+1}{M\nu}\frac{\gamma_k}{\gamma^{k+1}}.$$

Applying Lemma 1 the same way as before produces

$$\begin{aligned}\bar{R}_1(1-(1-\nu)^M)\nu &= -\frac{1}{M}\sum_{k=2}^{\infty}\frac{k+1}{M}\mathcal{O}(M^{-k/2})\\ &= \frac{1}{M}\sum_{k=2}^{\infty}\frac{k}{M}\mathcal{O}(M^{-k/2}) + \mathcal{O}(M^{-3})\\ &= \sum_{k=2}^{\infty}\mathcal{O}(kM^{-k/2-2}) + \mathcal{O}(M^{-3}).\end{aligned}$$

Put $f : k \mapsto k(M^{-\frac{k}{2}-2})$. Then

$$f'(k) = \frac{1}{2}M^{-\frac{k}{2}-2}\Big(2 - k\log(M)\Big);$$

i.e. $f'$ is negative for all $M \ge 3$. Because of this, $f$ is decreasing and its maximum is obtained when $k=2$ which leads to the conclusion

$$\sum_{k=2}^{\infty}\mathcal{O}(kM^{-k/2}) = \mathcal{O}(M^{-2/2-2}) = \mathcal{O}(M^{-3}).$$

For completeness, in the for us uninteresting case, when $M=2$ the maximum is obtained at $k=3$. As a consequence

$$\bar{R}_1 = \frac{1}{1-(1-\nu)^M}\mathcal{O}(M^{-3}),$$

which concludes the proof of the lemma. The result is in tune with the multiply cited preprint by Znidaric (2005) in which a different method of derivation is used.

**Lemma 3 (Expectations of $\rho$)** *Let $S_j$ and $S_j^+$ be zero truncated binomial random variables with probability of success $\nu_j$ and $\nu_j^+$ respectively. Let them have the same number of trials $M$. Define the ratio of them as $\rho_j = S_j^+ S_j^{-1}$. Then neither the expectation of the ratio $\rho_j$ nor the expectation of $\rho_j^2$ is increasing in $M$.*

*Proof* Consider the square of the ratio, use the tower property and then part 1 of Lemma 2 to get

$$\begin{aligned}\mathbb{E}[\rho_j^2] &= \mathbb{E}[(S_j^+ S_j^{-1})^2] = \mathbb{E}[S_j^{-2}\mathbb{E}[(S_j^+)^2 \mid S_j^+]]\\ &= \mathbb{E}[S_j^{-2}(1-(1-\nu_j^+)^M)^{-1}]\mathcal{O}(M^2)\\ &= \frac{1-(1-\nu_j)^M}{1-(1-\nu_j^+)^M}\mathcal{O}(M^{-2}M^2) = \frac{1-(1-\nu_j)^M}{1-(1-\nu_j^+)^M}\mathcal{O}(1)\\ &\le \frac{1}{1-(1-\nu_j^+)^M}\mathcal{O}(1),\end{aligned}$$

where we use that $\mathcal{O}(\mathbb{E}[(S_j^+)^2]) = \mathcal{O}(\mathbb{V}[S_j^+])$. Since the above expression is not increasing in $M$, neither is the expectation. By Lyapunov's inequality $\mathbb{E}[\rho_j] \le \sqrt{\mathbb{E}[\rho_j^2]}$. Thus, none of the expectations are increasing in $M$.

### *B.3.1 Proof of Convergence*

We seek the variance of $\pi^M(C_j)$. Define

$$S_j = \sum_{k=1}^{M}\mathbb{I}(\xi_k^{C_j} \in \Delta_{j-1}), \qquad X_j = \pi^M(C_j)S_j^+,$$

$$S_j^+ = \sum_{k=1}^{M}\mathbb{I}(\xi_k^{C_j} \in \Delta_j), \qquad \rho_j = \frac{S_j^+}{S_j},$$

and

$$T_j = \mathbb{E}[X_j^2], \qquad R_j = \mathbb{E}[\rho_j^2],$$
$$t_j = \mathbb{E}[X_j]^2, \qquad r_j = \mathbb{E}[\rho_j]^2.$$

Let $\nu_j$ and $\nu_j^+$ denote $\mathbb{P}(\xi_k^{C_j} \in \Delta_{j-1})$ and $\mathbb{P}(\xi_k^{C_j} \in \Delta_j)$ respectively. Note that $X_{j-1}$ is independent of $S_j$ and $S_j^+$. The variance of interest is

$$\begin{aligned}\mathbb{V}[\pi^M(C_j)] &= \mathbb{V}\left[\pi^M(C_{j-1})\frac{S_{j-1}^+}{S_j}\right] = \mathbb{V}\left[\frac{X_{j-1}}{S_j}\right] \\ &= \mathbb{E}[X_{j-1}^2]\mathbb{E}[S_j^{-2}] - \mathbb{E}[X_{j-1}]^2\mathbb{E}[S_j^{-1}]^2,\end{aligned}$$

where the product variance formula is used, see for instance Goodman (1960). The variables $S_j$ and $S_j^+$ are zero truncated binomial since the algorithm did not fail; thus, Lemma 2 implies that

$$\mathbb{V}[\pi^M(C_j)] = \frac{\left(1+\mathcal{O}(M^{-3})\right)\left(T_{j-1} - \frac{t_{j-1}}{1-(1-\nu_j)^M}\right)}{(M\nu_j)^2(1-(1-\nu_j)^M)}. \tag{18}$$

As $\mathbb{V}[X_j] = \mathbb{E}[X_j^2] - \mathbb{E}[X_j]^2 = T_j - t_j$, and on the other hand

$$\begin{aligned}\mathbb{V}[X_j] &= \mathbb{V}[\pi^M(C_j)S_j^+] = \mathbb{V}[X_{j-1}\rho_j] \\ &= \mathbb{E}[X_{j-1}^2]\mathbb{E}[\rho_j^2] - \mathbb{E}[X_{j-1}]^2\mathbb{E}[\rho_j]^2,\end{aligned}$$

we get the recursion $T_j - t_j = T_{j-1}R_j - t_{j-1}r_{j-1}$. Assume that the quantities $R_j$ and $r_j$ are known and consider the recurrence relation

$$\sqrt{t_j} = \mathbb{E}[X_j] = \mathbb{E}[\pi^M(C_j)S_j^+] = \mathbb{E}[X_{j-1}\rho_j] = \sqrt{t_{j-1}r_j}.$$

Since $t_1 = \mathbb{E}[X_1]^2 = \mathbb{E}[\pi^M(C_1)S_1^+]^2 = \mathbb{E}[1 \cdot S_1^+]^2 \triangleq r_1$ we obtain the solution $t_j = \prod_{s=1}^{j-1} r_s$ from Lemma 1. Furthermore, since $t_j$ is now a known constant we can write

$$T_j = T_{j-1}R_j \underbrace{-r_j(t_j + t_{j-1})}_{c_j} = T_{j-1}R_j + c_j. \tag{19}$$

Observe that $T_1 = \mathbb{E}[(\pi^M(C_j)1S_1^+)^2] = \mathbb{E}[(S_1^+)^2] \triangleq R_1$. The solution to (19) is given by Lemma 1 as

$$T_j = \prod_{s=1}^{j} R_s + \sum_{s=0}^{j-3}\left\{c_{j-s-1}\prod_{u=j-s}^{j} R_u\right\} + c_j.$$

The quantities $T_{j-1}$ and $t_{j-1}$ are linear combinations of $R_j$ and $r_j$ which are, by Lemma 3, not increasing in $M$. This, finally, implies that the expression for the variance in (18) will tend to zero as $M$ tends to infinity.

### B.4 Proof of Proposition 3

First observe that the expectation $\mathbb{E}[h(X)]$ can be written as

$$\sum_{j=1}^{W} \mathbb{E}[h(X) \mid X \in C_j \setminus \Delta_{j-1}]\mathbb{P}(X \in C_j \setminus \Delta_{j-1}). \tag{20}$$

Since $\pi^M(C_j \setminus \Delta_{1:j-1})$ is sampled from Algorithm 3 we have almost sure convergence to $\pi(C_j \setminus \Delta_{j-1})$. Further, since $\xi_k^{C_j} \overset{d}{=} X \mid X \in C_j$ we have by the strong law of large numbers, or in the case when $\xi^{C_j}$ is a $\pi(\cdot \mid C_j)$-irreducible MH chain by Theorem 7.4 in Robert and Casella (2004), that the sum $\frac{1}{M}\sum_{k=1}^{M} h(\xi_k^{C_j})$ converges almost surely to $\mathbb{E}[h(X) \mid X \in C_j]$ as $M$ tends to infinity. By continuity we have that the product of these two estimates almost surely converges to the product of their limits as $M$ tends to infinity

$$\begin{aligned}&\frac{1}{M}\sum_{k=1}^{M} h(\xi_k^{C_j})\pi^M(C_j \setminus \Delta_{1:j-1}) \\ &\quad \longrightarrow \mathbb{E}[h(X) \mid X \in C_j]\pi(C_j \setminus \Delta_{1:j-1}).\end{aligned}$$

Combining this convergence with the fact that we for each $C_j$ discard the samples that are in $\Delta_{j-1}$, i.e. the $\xi$-samples we allow to contribute to $\eta$ are distributed as $X \mid X \in C_j \setminus \Delta_{j-1}$, yields almost surely for $\eta$

$$\frac{1}{M}\sum_{k=1}^{M} \eta_k \xrightarrow[M\to\infty]{} \sum_{j=1}^{W} \mathbb{E}[h(X) \mid X \in C_j \setminus \Delta_j]\mathbb{P}(X \in C_j \setminus \Delta_j),$$

where we recall that $\pi(C_j \setminus \Delta_j) \triangleq \mathbb{P}(X \in C_j \setminus \Delta_j)$. The limit above coincides with (20) which is what we wanted to establish.

## C Details on Implementation of the Stochastic Volatility Model

In this Section details on the implementaion of the stochastic volatility model are given. Recall that the volatility model is for the *log returns*, defined as $Y_k = \log(\frac{S_k}{S_{k-1}})$ and that the model for the log returns is the partially observed hidden Markov model

$$\begin{aligned} Y_k &= \beta e^{X_k/2} u_k, \\ X_k &= \phi X_{k-1} + \sigma w_k, \end{aligned} \tag{21}$$

where $Y_k$ are the observed log returns and $X$ is the hidden process driving $Y$.

The variables $u_k$ and $w_k$ are jointly normally distributed with zero mean, unit variance and correlation $\rho$. The object is to estimate the four parameters $\phi, \beta, \rho$ and $\sigma$. This will be done with a Bayesian approach. Since $X$ is not observed sequential Monte Carlo methods will be employed to obtain estimates of it. More specifically, we will use the PMMH sampler, see Andrieu et al (2010), or Olsson and Rydén (2011) for an extension. The example is taken from Hallgren (2011) where the posteriors and choice of priors can be found. The model in (21) is a variation of one introduced by Taylor (1982), where $u$ and $w$ has zero correlation.

For more on the history of the model and an economic interpretation, see Shephard (2005). In Hallgren (2011) the model is evaluated on market data, the results indicate that it performs better (with respect to value at risk and expected shortfall) than the model without correlation. In this paper we will however confine ourselves to fitting the parameters. The variable of interest is $Z = (\phi, \beta, \rho, \sigma, X_{0:T})$. We are not directly interested in $X$ for the purpose of calibration but the PMMH sampler requires us to sample $X$-trajectories. It is natural to let $\sigma$ be greater than zero and to let the modulus of $\rho$ be smaller than 1. Note that switching the sign of $\beta$ is the same thing as switching the sign of $u$, which in turn implies that we switch the sign of $\rho$. Thus it is natural to keep $\beta$ positive (or negative). Further view $X$ as an autoregressive volatility process. This combined with empirical studies, found in Hallgren (2011), engenders a positive $\phi$; to maintain stationarity it is forced to be smaller than

1. The restrictions on the variables produce the state space for $Z$, $\mathsf{S}^Z = \{[0,1), \mathsf{R}^+, [-1,1], \mathsf{R}^+, \mathsf{R}^T\}$.

First observe that, under some assumptions, Theorem 4 in Andrieu et al (2010) implies that we are dealing with a MH chain, which we can parallelize in our setting. In order to apply the Decomposition sampling, Algorithm 1, we need to divide the space $\mathsf{S}^Z$ into subsets. We choose a simple split, i.e. divide the space into two parts, $C_1$ and $C_2$. The parts are chosen in a heuristic way. In Section 2.3 we proposed a more sophisticated method for selecting the splits. The space is divided in the $\phi$-variable in the following way

$$\begin{aligned} C_1 &= \{[0, 0.55+0.01], \mathsf{R}^+, [-1,1], \mathsf{R}^+, \mathsf{R}^{T+1}\}, \\ C_2 &= \{[0.55-0.01, 1), \mathsf{R}^+, [-1,1], \mathsf{R}^+, \mathsf{R}^{T+1}\}. \end{aligned} \tag{22}$$

Thus, the intersection $\Delta$ is given by

$$\Delta = \{[0.54, 0.56], \mathsf{R}^+, [-1,1], \mathsf{R}^+, \mathsf{R}^{T+1}\}.$$

This is used in the Decomposition sampling.